\documentclass[conference]{IEEEtran}
\IEEEoverridecommandlockouts
\usepackage{cite}
\usepackage{amsmath,amssymb,amsfonts}
\usepackage{graphicx}
\usepackage{xcolor}
\usepackage{array}
\usepackage{url}
\usepackage{booktabs}
\usepackage{colortbl}  
\definecolor{tablegray}{gray}{0.80}  
\definecolor{anchoracc}{HTML}{36648B}\definecolor{anchoroff}{HTML}{D55E00}\definecolor{anchorcaller}{HTML}{E69F00}\definecolor{anchorreturn}{HTML}{009E73}

\newcommand{\sig}[1]{{\boldmath\bfseries #1}$^{*}$}
\newcommand{\sigp}[1]{{\boldmath\bfseries #1}}
\newcommand{\good}[1]{\underline{{\boldmath\bfseries #1}}$^{*}$}
\newcommand{\goodci}[3]{{\boldmath\bfseries\begin{tabular}[t]{@{}r@{}}\underline{#1}$^{*}$\\(#2, #3)\end{tabular}}}
\newcommand{\ci}[3]{\begin{tabular}[t]{@{}r@{}}#1\\(#2, #3)\end{tabular}}
\newcommand{\sigci}[3]{{\boldmath\bfseries\begin{tabular}[t]{@{}r@{}}#1$^{*}$\\(#2, #3)\end{tabular}}}
\newcommand{\hd}[2]{\begin{tabular}[b]{@{}c@{}}#1\\#2\end{tabular}}
\newcommand{\cii}[3]{#1 (#2, #3)}
\newcommand{\sigcii}[3]{{\boldmath\bfseries #1$^{*}$ (#2, #3)}}
\newcommand{\goodcii}[3]{{\boldmath\bfseries \underline{#1}$^{*}$ (#2, #3)}}
\newsavebox{\fitbox}
\makeatletter
\newenvironment{fittable}{\begin{lrbox}{\fitbox}}{\end{lrbox}%
  \typeout{FITTABLE line \the\inputlineno: natural \the\wd\fitbox\space line \the\linewidth\space font \f@size pt}%
  \ifdim\wd\fitbox>\linewidth\resizebox{\linewidth}{!}{\usebox{\fitbox}}\else\usebox{\fitbox}\fi}
\makeatother
\usepackage{adjustbox}
\usepackage{fvextra}  
\usepackage{tikz}
\usetikzlibrary{arrows.meta,positioning,fit,backgrounds}
\graphicspath{{plots/}}
\usepackage[hidelinks]{hyperref}

\begin{document}

\title{Code-Anchored Review Feedback Improves Correctness and Lowers Hosted Editing Cost on Large R Files}

\author{\IEEEauthorblockN{William Franz Lamberti}
\IEEEauthorblockA{
\textit{Computational and Data Sciences}\\
\textit{College of Science}\\
\textit{George Mason University}\\
Fairfax, VA, United States \\
wlamber2@gmu.edu
}
}

\maketitle

\begin{abstract}
Tokens drive the cost and latency of generative AI (GAI) code editing, and correctness decides whether an edit is useful. We study how the format of a reviewer's feedback, a human-AI collaboration (HAIC) channel, impacts both. In a paired experiment on 36 synthetic base-R tasks with seven models, the same requested changes were delivered as a holistic list (control) or as the export of FileMark (treatment). That code-review extension anchors each comment to the code it concerns and quotes it. For the typical task, total tokens changed by $-4.2\%$ to $+14.7\%$ across models, while output tokens fell by 14.7\% (Claude Opus) and 50.5\% (Claude Sonnet). Since an output token costs five times as much as an input token, the hosted cost of an edit on files over 100 lines fell by 20.0\% (Opus) and 46.3\% (Sonnet). 
The treatment raised the correct-run share of four of five local models by 24\% to 46\%. A decomposition experiment places these gains in the format, not the wording, and mostly in the quoted code. Misplaced anchors cut correctness sharply. 
Code-anchored feedback thus lowers hosted editing cost on large R files and makes weaker models more correct, provided the anchors are accurate.
\end{abstract}

\section{Introduction}

Editing and repairing source code is still a human-AI collaboration (HAIC): generative artificial intelligence (GAI) systems propose and apply the changes while developers direct and review them \cite{liang_survey_2024}. Utilizing the words and definitions from others, we define artificial intelligence (AI) as a computer's ability to ape human cognitive functions such as learning and problem solving \cite{lamberti_ai_policy_2024}. GAI is defined as the specialized branch of AI that creates content with a spatial component (i.e., text, images, or code) from user-provided prompts after training on large corpora of existing content \cite{lamberti_ai_policy_2024}. Every such interaction with a GAI editor has a price. The tokens an exchange consumes, as input and as output, drive its inference cost, latency, and energy consumption \cite{schwartz_green_2020, luccioni_power_2024}. The price of a hosted exchange is well understood: providers charge a price per input token for the prompt and a higher price per output token for the answer, so the price follows from the two token counts \cite{chen_frugalgpt_2024, anthropic_pricing_2026}. The output tokens weigh most: a model generates them one at a time, so they set the time a call takes \cite{kwon_pagedattention_2023}, and they are priced several times higher than input tokens \cite{anthropic_pricing_2026}. How the delivery of feedback changes those counts, however, is not well understood.

A natural source of well-structured feedback for HAIC already exists in developer practice: code review. Modern code review is localized.  For instance, reviewers attach specific comments to specific lines, and understanding the code under review is their main challenge \cite{bacchelli_expectations_2013}. However, GAI assistants are typically given feedback in a different form, as a single holistic prompt. FileMark \cite{lamberti_filemark_2026} brings this localized form to GAI workflows. It is an extension for VS Code and VSCodium that records pull-request-style inline comments on any file, without requiring a repository or pull request, and exports them to a structured \texttt{.feedback.md} artifact intended to be handed to a GAI assistant. In collaboration terms, it makes the human half of the HAIC editing loop machine-readable: the reviewer's judgment about what to change and where, recorded in the localized form code review already uses.

On the model side, whether a GAI assistant can act on such feedback is a question of automated program repair. Large pre-trained language models are strong program repairers, outperforming dedicated repair techniques when prompted directly \cite{xia_automated_2023}. Benchmarks such as SWE-bench evaluate models on repository-scale issue resolution \cite{jimenez_swebench_2024}, and EvalPlus shows that rigorous, test-augmented evaluation frequently overturns optimistic correctness claims \cite{liu_is_2023}. We adopt the same execution-based standard where every generated script is run in a sandbox and validated against ground truth recomputed from the executed data.  However, our object of study is the feedback format rather than the model.

Localized review comments are established practice \cite{bacchelli_expectations_2013}, and subsequent work automated parts of the review loop with pre-trained models, learning to generate review comments and to revise code in response to them \cite{tufano_using_2022, li_automating_2022}. Our work inverts this direction: rather than generating review comments, we ask whether human-authored comments anchored to the code they concern, exported in a structured machine-readable format, make a GAI editor more effective than the same feedback expressed holistically.

How feedback is phrased matters as much as what it contains: prompt formulation is known to change model behavior \cite{wei_chainofthought_2022}. Further, studies of GAI pair programming, itself a HAIC practice, find that programmers struggle when they must locate and repair the misalignment between intent and generated code \cite{vaithilingam_expectation_2022, barke_grounded_2023}. Feedback anchored to the code is a natural way to communicate exactly that misalignment. To our knowledge, no prior controlled experiment separates the format of review feedback from its wording.

The stakes of the efficiency question are equally well established. The energy and monetary costs of deep learning have been quantified for training \cite{strubell_energy_2019}. More recently, the energy of inference has been measured directly, and for widely used systems it comes to dominate the lifetime cost \cite{luccioni_power_2024}. Green AI argues that efficiency should be a first-class evaluation criterion alongside accuracy \cite{schwartz_green_2020}. Token efficiency is a co-primary endpoint of this study, not an afterthought. Finally, our fully scripted, seeded, and versioned pipeline responds to persistent reproducibility gaps in AI research \cite{gundersen_state_2018}.

Against this backdrop, our experiment asks four questions about this form of HAIC: \begin{enumerate}
  \item Holding the requested changes constant, does delivering them as structured, line-anchored feedback yield better outcomes than an equivalent holistic prompt?
  \item Does the answer depend on the capability of the model and on the size of the file being edited?
  \item Which part of the feedback carries the effect, its wording or its format? 
  \item What happens when an anchor points to the wrong code?
\end{enumerate}
The first two questions are answered by the main experiment (Sections~\ref{sec:design_main} and~\ref{sec:results_main}) and the last two by the revision experiments that referees' questions prompted (Sections~\ref{sec:newarms} and~\ref{sec:newresults}).

These questions concern HAIC. The anchored arm carries a reviewer's localized judgment in the form modern code review already produces.  Thus, the experiment quantifies what that human review attention is worth to an AI editing workflow, in tokens and in validated correctness.

We find that code-anchored feedback lengthens the request and shortens the Claude models' answers on large files. Total tokens therefore rise on small files for nearly every model and fall on large files for the Claude models and gemma4, and the hosted cost of an edit falls on files over 100 lines. Code-anchored feedback also makes four of five local models correct far more often.  A decomposition experiment shows that the anchored format, and above all the quoted code, carries the effect, not the wording of the comments. A misplaced-anchor experiment shows the other side: a model follows a wrong anchor, and its correctness collapses.  All tasks, harness code, raw model outputs, analyses, and the record of every correction are released for reproduction.

\section{Experimental Design}
\label{sec:methods}

This section describes the \textbf{main experiment} (Section~\ref{sec:design_main}) and then the three \textbf{revision experiments} that the referees' questions prompted (Section~\ref{sec:newarms}). Section~\ref{sec:results} reports their results in the same order, and Appendix~\ref{app:timeline} gives the timeline of every run and analysis.

\subsection{Main Experiment}
\label{sec:design_main}

We construct a fully paired design in which both arms are rendered from one change list, so they request the same changes in the same order and number. The arms differ in how the changes are delivered, in wording and in format (Section~\ref{sec:conditions}). Each run is thus a minimal HAIC exchange: a reviewer's change list on the human side, a model on the AI side, and between them the delivery channel under study. We evaluate five local and two frontier models on a tiered, template-generated benchmark of 36 model-building \textbf{tasks}. A task is one flawed base-R script, the list of changes a reviewer requests, and a validator that checks the result (Section~\ref{sec:tasks}). A \textbf{run} is one model's attempt at one task under one arm, and a \textbf{replicate} repeats every run with a new seed. The tasks form six file-size tiers, from under 10 to over 1{,}000 lines. On large files the model returns \texttt{SEARCH}/\texttt{REPLACE} edit blocks of its own choosing. We measure two co-primary outcomes. The first is \textbf{token efficiency}: the total tokens an exchange consumes, request and answer together, and, for the hosted models, the price of the call. Within the total, the output tokens the model generates are the component that generation time and output pricing scale with. The second is \textbf{correctness}, whether the model accurately implements the requested changes.

We call this comparison of the control and treatment arms on all 36 tasks, with 13 replicates for every task, arm, and model, the \textbf{main experiment} (6{,}552 runs, made from June 25 to July 7, 2026). Four further experiments complement it. An exploratory function-patch wave from the same period changes how the large-tier edits are applied (Appendix~\ref{app:patch}). Three revision experiments with new arms on subsets of the tasks separate the wording of the feedback from its format, test misplaced anchors, and separate the output contract (Section~\ref{sec:contract}) from file size (Sections~\ref{sec:newarms} and~\ref{sec:newresults}). 

\subsubsection{Conditions and Matched Requests}
\label{sec:conditions}
Each task provides a flawed or incomplete base-R script (\texttt{seed.R}) and a canonical list of requested HAIC changes (\texttt{changes.json}). Both arms are derived from this list deterministically (Fig.~\ref{fig:workflow}). The \textbf{treatment} input is a FileMark \texttt{feedback.md} export. Each change becomes a comment anchored to its line span that quotes the selected source text, exactly as the extension produces it. The \textbf{control} input is a holistic plain-text list of the same changes. Figure~\ref{fig:prompts} shows both inputs for one task as they were sent. Appendix~\ref{app:prompts} prints a complete prompt pair and counts the localization cues.

The arms are matched in what they request: the changes, in the same order and number, inside an identical prompt. Every stored prompt (9{,}828 of 9{,}828) is byte-identical to the prompt rebuilt from its task files. However, the arms are not matched word for word. Each change carries two texts. The control text is a short goal statement (55 characters on average). The treatment text adds a diagnosis and, in 112 of the 129 items, the exact replacement code (90 characters on average). No control text names a line, a function, or a block, and none quotes code. Most do name script variables (115 of 129), and half describe the error by contrast, as in ``mean, not median'' (65 of 129). The treatment is therefore the FileMark-style representation as a whole: wording, line anchor, quoted span, and layout together.

The design is \textbf{paired}: every task instance runs under both arms for every model and replicate. In HAIC terms, the two arms deliver the same reviewer request through two channels: anchored to its lines as code review produces it, or written as a holistic prompt.

\begin{figure*}[!t]
  \centering
  \begin{minipage}[t]{0.45\linewidth}
    \footnotesize\textbf{(a) Original code:} the four flawed lines of \texttt{seed.R} (seed line numbers)
    \begin{Verbatim}[fontsize=\scriptsize, frame=single, breaklines]
  36    gm <- tapply(d$y, d$g, median)       # [block 3]
  68    mu <- mean(d$x3)                     # [block 7]
 131    cnt <- sum(d$x1 > 0)                 # [block 15]
 138    m <- lm(y ~ x1, data = d)            # [block 16]
    \end{Verbatim}
  \end{minipage}\hfill
  \begin{minipage}[t]{0.53\linewidth}
    \footnotesize\textbf{(b) Control, prompt lines 201--207}
    \begin{Verbatim}[fontsize=\scriptsize, frame=single, breaklines]
Please make the following changes to the script:

1. Compute group means of y by g (mean, not median).
2. Compute the mean of x3 with na.rm = TRUE so it is not NA.
3. Count observations with x1 > 1 (not > 0).
4. Include x2 in the linear model (y ~ x1 + x2) for this block.

    \end{Verbatim}
  \end{minipage}

  \medskip
  \begin{minipage}[t]{0.49\linewidth}
    \footnotesize\textbf{(c) Treatment, prompt lines 201--228}
    \begin{Verbatim}[fontsize=\scriptsize, frame=single, breaklines, breakafter=/, breakaftersymbolpre={}]
# Feedback: seed.R
File: `/home/billy/Documents/Projects/research/filemark_usage/tasks/t4_lib_01/seed.R`
Exported: deterministic (generated)

---

## Comment 1 (line 36)

**Selected text:**
```
gm <- tapply(d$y, d$g, median)       # [block 3]
```

**Feedback:** This should report group MEANS, not medians. Use tapply(d$y, d$g, mean).

---

## Comment 2 (line 68)

**Selected text:**
```
mu <- mean(d$x3)                     # [block 7]
```

**Feedback:** x3 contains NA, so mean(d$x3) is NA. Use mean(d$x3, na.rm = TRUE) so mu is finite.

---

    \end{Verbatim}
  \end{minipage}\hfill
  \begin{minipage}[t]{0.49\linewidth}
    \footnotesize\textbf{(c, continued) lines 229--251}
    \begin{Verbatim}[fontsize=\scriptsize, frame=single, breaklines, breakafter=/, breakaftersymbolpre={}]
## Comment 3 (line 131)

**Selected text:**
```
cnt <- sum(d$x1 > 0)                 # [block 15]
```

**Feedback:** The threshold is wrong: count x1 > 1 (not x1 > 0). Use sum(d$x1 > 1).

---

## Comment 4 (line 138)

**Selected text:**
```
m <- lm(y ~ x1, data = d)            # [block 16]
```

**Feedback:** This model omits predictor x2. Change it to lm(y ~ x1 + x2, data = d) so the fit uses both predictors.

---


    \end{Verbatim}
  \end{minipage}
  \caption{Both prompts for task \texttt{t4\_lib\_01} (edit-block contract), exactly as sent, with long lines wrapped for printing at a hooked arrow ($\hookrightarrow$) that is not part of the prompt. Prompt lines 1--200 (instructions, answer format, and the full 179-line script) are byte-identical in both arms and for every model, so panels (b) and (c) are the whole difference.}
  \label{fig:prompts}
\end{figure*}

\subsubsection{Task Bank and Size Tiers}
\label{sec:tasks}
Tasks are generated by parameterized templates and committed for inspection. Small tasks are multi-issue composites covering common model-building work in base R \cite{rcore_r_2025}: linear and generalized linear modeling, train/test evaluation, missing-data handling, regression diagnostics, hand-coded $k$-fold cross-validation, and group aggregation with base graphics. Larger tasks are produced by a ``library'' generator that composes a coherent program from $K$ named analysis functions plus a driver, the code at the end of the script that calls each function in turn (Fig.~\ref{fig:generator}). The number $K$ sets program length and places the 30 generator-built instances across tiers T2--T6 by lines of code. Together with the six composite T1 tasks, each of the 36 instances lands in one of six \textbf{size tiers} (Table~\ref{tab:tiers}). To reach the tier minima, the generator adds inert comment lines to the header of some seeds (Appendix~\ref{app:padding}). T6 crosses its 1{,}000-line threshold only with this padding: each T6 seed carries 136 padding lines (13.5\%), and its 874 lines of code alone would fall in the T5 range. Each task carries a small, fixed set of injected flaws (two to five, set per template and growing only mildly with tier).  Thus, larger files chiefly make flaws harder to locate.  This is the regime where line anchoring should matter most. Injected flaws include omitted predictors, missing \texttt{na.rm} handling, wrong summary statistics, wrong standardization denominators, and wrong count thresholds.

\begin{figure*}[!t]
  \centering
  \begin{adjustbox}{max width=\textwidth}
\begin{tikzpicture}[
    font=\small\bfseries,
    >={Stealth[length=2.4mm]},
    node distance=3mm and 6mm,
    box/.style={draw, line width=0.5pt, rounded corners=2pt, align=center,
                inner sep=3pt, minimum height=9mm, fill=gray!8, text=black,
                text width=30mm},
    arr/.style={->, line width=1pt, black!70}
  ]
  \node[box] (params) {template\\parameters: $K$\\functions, seed,\\flaw list};
  \node[box, right=of params, fill=orange!8] (gen) {library generator\\($K$ sets length,\\tiers T2--T6)};
  \node[box, right=of gen, fill=green!8, text width=34mm] (bank) {36 task instances,\\tiers T1--T6\\(Table~\ref{tab:tiers})};
  \node[box, below=3mm of gen, fill=gray!12] (t1) {six composite\\templates (T1)};
  \draw[arr] (params) -- (gen);
  \draw[arr] (gen) -- (bank);
  \draw[arr] (t1.east) -| (bank.south);
\end{tikzpicture}\end{adjustbox}
  \caption{Task-bank construction. A library generator composes $K$ named analysis functions plus a driver, with $K$ setting the length of tiers T2--T6, and six composite templates provide T1, for the 36 tasks of Table~\ref{tab:tiers}.}
  \label{fig:generator}
\end{figure*}

\begin{table*}[!t]
\caption{Task bank: size tiers, task counts, and the output contract, which is the same in both arms (control and treatment) of every pair. A gray background marks the large tiers T4--T6, here and in every later table.}
\label{tab:tiers}
\centering
\begin{fittable}
\begin{tabular}{lccl}
\toprule
Tier & Lines of code & Tasks & Output contract (both arms) \\
\midrule
T1 & $<$10 & 6 & full rewrite \\
T2 & 10--50 & 6 & full rewrite \\
T3 & 50--100 & 6 & full rewrite \\
\rowcolor{tablegray} T4 & 100--500 & 6 & edit blocks (\texttt{SEARCH}/\texttt{REPLACE}) \\
\rowcolor{tablegray} T5 & 500--1{,}000 & 6 & edit blocks (\texttt{SEARCH}/\texttt{REPLACE}) \\
\rowcolor{tablegray} T6 & $\geq$1{,}000 & 6 & edit blocks (\texttt{SEARCH}/\texttt{REPLACE}) \\
\bottomrule
\end{tabular}
\end{fittable}
\end{table*}

\subsubsection{Output Contract}
\label{sec:contract}
The \textbf{output contract} is the form in which the model is instructed to return its answer. The contract depends on tier and is identical across arms within a tier. For small tiers (T1--T3) the model returns the entire corrected script. For large tiers (T4--T6) the model returns one or more \texttt{SEARCH}/\texttt{REPLACE} edit blocks of its own choosing, the ``diff'' edit format of the Aider coding assistant \cite{aider_editformats}. Each block consists of a span of lines copied verbatim from the current script and the lines that should replace it. The harness applies each block by exact, unique, fixed-string match and validates the resulting full program. The model chooses what to touch and at what granularity. A response without an applicable block is recorded as an execution error. The complete contract grammar and applier semantics are documented in the repository.

This large-tier contract mirrors how FileMark exports are consumed in practice: the developer hands the \texttt{feedback.md} artifact to a coding assistant, and the assistant decides which edits to make and applies them itself. Regenerating an entire 1{,}000-line file per request is not how such files are edited in practice, and pilot runs showed local models degenerating when asked to do so. An exploratory contract in which the model instead returns whole corrected functions and our harness splices them into the file by name lifts the mechanical application burden from the model. This is reported in full in Appendix~\ref{app:patch}. The contract thus fixes the division of labor inside the collaboration: the human contributes anchored judgment, the assistant chooses and applies the edits, and the exploratory variant shifts only the mechanical application step from the AI collaborator to the tooling.

Two design properties follow. First, because the contract is fixed within a tier and identical across arms, it cannot confound the treatment effect: every control--treatment pair shares its task, tier, and contract. Second, the contract co-varies with tier: T1--T3 use full rewrites and T4--T6 use edit blocks. Any difference between tiers, including a change in the treatment effect from one tier to another, therefore mixes file size with contract. A contract-control experiment (Section~\ref{sec:newarms}) separates the two by running the edit-block contract on T1--T3 and full rewrites on T4.

\subsubsection{Models}
\label{sec:models}
We evaluate five local open-weight models and two hosted frontier models (Table~\ref{tab:models}). The local models ran through Ollama's HTTP interface at temperature 0.4. Each replicate number served as the random seed, so both arms of a pair share a seed. The hosted models ran through the Claude Code command-line interface (\texttt{claude -p}), invoked by alias from the repository directory. Each hosted call therefore also carried Claude Code's own system prompt and tool definitions and the author's \texttt{CLAUDE.md} instruction files. This context was the same in both arms, and it accounts for most of a hosted call's input tokens. In 11 of the 2{,}808 hosted runs the model also used a tool (two to five turns).

The interface records the resolved model in every response. An audit of all archived responses confirms that every analyzed Opus run used \texttt{claude-opus-4-8} and every Sonnet run used \texttt{claude-sonnet-5}. For 142 early runs the alias had resolved to \texttt{claude-sonnet-4-6}. These runs were re-executed on \texttt{claude-sonnet-5} under the same run identifiers: 71 in the main experiment and 71 in the exploratory function-patch wave. The superseded responses are kept in the repository and are not counted in any total.

\begin{table*}[!t]
\caption{Models under study.}
\label{tab:models}
\centering
\begin{fittable}
\begin{tabular}{llll}
\toprule
Model & Parameters & Class & Access \\
\midrule
\texttt{llama3.1} & 8B & general & local (Ollama) \\
\texttt{qwen2.5-coder:14b} & 14B & code & local (Ollama) \\
\texttt{gemma4:26b} & 26B & general & local (Ollama) \\
\texttt{gemma3:27b} & 27B & general & local (Ollama) \\
\texttt{qwen3-coder:30b} & 30B & code & local (Ollama) \\
\addlinespace
Claude Sonnet (\texttt{claude-sonnet-5}) & undisclosed & frontier & hosted (Claude CLI) \\
Claude Opus (\texttt{claude-opus-4-8}) & undisclosed & frontier & hosted (Claude CLI) \\
\bottomrule
\end{tabular}
\end{fittable}
\end{table*}

\subsubsection{Outcomes}
\label{sec:outcomes}
Each generated script runs in a sandbox, a separate R process with a 30-second time limit (Appendix~\ref{app:sandbox}), and is graded by a task-specific validator that recomputes expected values from the executed objects. For example, a stored adjusted $R^2$ must match \texttt{summary(model)} to within $10^{-6}$, and a required plot file must exist and be non-empty. Fig.~\ref{fig:scoring} summarizes the scoring pipeline. We record eight outcomes.
\begin{enumerate}
  \item The \textbf{partial-credit score} is the fraction of a task's graded checks passed.
  \item Binary \textbf{correctness} means that all checks passed.
  \item The three-level \textbf{run status} is \texttt{error} (the returned edit could not be applied, or the program does not run), \texttt{runs\_fail} (it runs but fails validation), or \texttt{correct}.
  \item \textbf{Total tokens} are the tokens the exchange consumes: the task prompt plus the output tokens (Table~\ref{tab:costparts}). They are the primary efficiency outcome. The prompts are identical across models, so the task prompt is counted once for every model, with the tokenizer of \texttt{qwen3-coder:30b}. The \emph{total as run} counts the input each model actually received, with its own tokenizer, and for the hosted models it includes Claude Code's own context (Section~\ref{sec:models}).
  \item \textbf{Output tokens} are the tokens the model generated, including any hidden reasoning a provider bills as output. They are the part of the total that the model controls, they are unaffected by prompt caching, and they are comparable across providers.
  \item \textbf{Cost}, for the hosted models, in two views (Table~\ref{tab:costparts}). The \emph{cost as run} is the price in US dollars that the command-line interface estimated for the call. The \emph{cost via API} is the price the same exchange would have had if sent directly to the API. In other words, it is the task prompt's tokens at the input price plus the output tokens at the output price \cite{chen_frugalgpt_2024}. Both views use the prices behind the interface's estimates. Per million input and output tokens, these are \$5 and \$25 for Opus, its list prices, and \$3 and \$15 for Sonnet, 1.5 times its list price for every token class \cite{anthropic_pricing_2026}. A common price factor cancels in every percentage change, so no result depends on it. An output token thus costs five times as much as an input token.
  \item \textbf{Constraint adherence} means base R only, with no external packages.
  \item \textbf{Collateral damage} means that an edit changed the result of a block the reviewer did not ask to change. The validators check only the flawed blocks, so the revision adds this check: the result of every unflawed block is compared with the seed's. By design, a T2 seed has only the two blocks that its change list flaws, so the check has nothing to compare there and applies to T3--T6. The check finds planted damage in all 942 unflawed blocks of the task bank, and none in reference fixes.
\end{enumerate}
Token counts, cost, and the time per call are analyzed on the natural logarithm scale, $\log(\text{count} + 1)$ for tokens, because they are right-skewed and the treatment changes them by a factor rather than by a fixed amount. On the raw scale the run-level output tokens of six of the seven models have a skewness of 0.73 to 7.64, and on the log scale every model's lies between $-1.09$ and 0.79. Fig.~\ref{fig:hier_sonnet} in Appendix~\ref{app:hier} shows them before and after the transformation for Claude Sonnet, as a representative model, and the repository draws them for every model. Effects are back-transformed to percentage changes (Section~\ref{sec:power}). 

The prompt asks for the answer in a single fenced code block, and the scorer read only the first block of each response. Some responses placed each edit in a block of its own, so their later edits were never applied. We therefore report two scorings. The \emph{strict} scoring is the original rule. The \emph{lenient} scoring reads every edit block and every function definition in the whole response and applies them with the same exact matching. It was pre-registered before it was computed, and a hand audit of 40 changed runs found every recovery genuine. A correctness conclusion needs evidence under both scorings.

\begin{table*}[!t]
\caption{Main experiment: what each token and cost measure includes (X: included). Totals count the task prompt with one tokenizer, totals as run count the input each model read, and cached input is priced at twice (writes) or a tenth (reads) of the input price.}
\label{tab:costparts}
\centering
\begin{fittable}
\begin{tabular}{lcccc}
\toprule
 & \multicolumn{2}{c}{Tokens} & \multicolumn{2}{c}{Cost (hosted models)} \\
\cmidrule(lr){2-3}\cmidrule(lr){4-5}
Component & Total & Total as run & Via API & As run \\
\midrule
Task prompt: instructions, answer format, and script & X & X & X & X \\
Task prompt: requested changes (differs between arms) & X & X & X & X \\
Output tokens, including hidden reasoning & X & X & X & X \\
\addlinespace
Chat template, counted with each local model's tokenizer & & X & & \\
Claude Code's system prompt and tool definitions & & X & & X \\
Prompt-cache prices for cached input & & & & X \\
The interface's side call to a second model & & & & X \\
\midrule
Models & all & all & hosted & hosted \\
\bottomrule
\end{tabular}
\end{fittable}
\end{table*}

\begin{figure*}[!t]
  \centering
  \begin{adjustbox}{max width=\textwidth}
\begin{tikzpicture}[
    font=\small\bfseries,
    >={Stealth[length=2.4mm]},
    node distance=3.5mm and 4mm,
    box/.style={draw, line width=0.5pt, rounded corners=2pt, align=center,
                inner sep=3.5pt, minimum height=8mm, fill=gray!8, text=black,
                text width=34mm},
    bad/.style={box, fill=red!12, text width=30mm},
    good/.style={box, fill=green!12, text width=30mm},
    arr/.style={->, line width=1pt, black!70}
  ]
  \node[box, text width=44mm] (out) {model output (either arm)};
  \node[box, below left=4mm and -14mm of out] (small) {T1--T3: returned\\script \emph{is} the candidate};
  \node[box, below right=4mm and -14mm of out] (large) {T4--T6: edit blocks\\applied to the file};
  \node[box, below=26mm of out] (run) {sandboxed\\\texttt{Rscript} run};
  \node[bad, right=of run] (rerr) {edit fails to apply\\or run throws\\$\Rightarrow$ \texttt{error}};
  \node[box, below=of run] (val) {validator's\\graded checks};
  \node[good, below left=3.5mm and -10mm of val] (score) {score $= n_{\text{pass}}/n_{\text{total}}$};
  \node[good, below right=3.5mm and -10mm of val] (corr) {all pass $\Rightarrow$\\\texttt{correct}};
  \draw[arr] (out) -- (small);
  \draw[arr] (out) -- (large);
  \draw[arr] (small) -- (run);
  \draw[arr] (large) -- (run);
  \draw[arr] (run) -- (rerr);
  \draw[arr] (run) -- (val);
  \draw[arr] (val) -- (score);
  \draw[arr] (val) -- (corr);
\end{tikzpicture}\end{adjustbox}
  \caption{Scoring pipeline. Every run yields a three-level run status, a partial-credit score, and binary correctness; both arms pass through the identical pipeline.}
  \label{fig:scoring}
\end{figure*}

\subsubsection{Experimental Environment}
\label{sec:env}
All local runs executed on a single workstation: Ubuntu 20.04.6 LTS, an Intel Core i7-12700K CPU, 64\,GB of RAM, and an NVIDIA GeForce RTX 3090~Ti GPU with 24\,GB of memory. Ollama 0.20.4 served the local models with its default quantized weights. The original runs did not set Ollama's context window, so it stayed at the default of 4{,}096 tokens. Every T5 and T6 prompt is longer than that (about 5{,}300 to 12{,}000 tokens), and Ollama kept only the end of each one. Those runs therefore never saw the instructions or most of the script.
On T1 to T4 the whole prompt fit, but the window must also hold the answer, so 637 long answers were cut off. Of these, 434 were gemma4's, in every tier, 156 were gemma3's on T4, and 47 were other models' on T4 (Appendix~\ref{app:truncation}). 

We re-ran every affected local cell with a 32{,}768-token window and otherwise identical settings: 4{,}836 of the 7{,}020 local runs, always both arms and all replicates of a cell. Before combining re-run and original cells, we confirmed that the model weights and runtime were unchanged and that original runs reproduce byte-for-byte under the original settings (10 of 10). All local results below use the corrected runs. The truncated originals are kept, and Appendix~\ref{app:truncation} documents the problem. The hosted models were reached over the network. All orchestration, validation, and analysis used R 4.5.2 \cite{rcore_r_2025}.

\subsubsection{Design and Analysis}
\label{sec:power}
The main experiment ran 13 replicates of all 36 tasks under both arms: 468 pairs per model and 6{,}552 runs. The original runs executed from June 25 to July 7, 2026. The small tiers, which both output contracts share, and the exploratory function-patch wave ran first, and the edit-block wave of the main experiment ran on July 6--7. The main experiment uses the edit-block contract on its large tiers because that contract mirrors how coding assistants consume FileMark exports (Section~\ref{sec:contract}). The revision's re-runs and new arms ran from October 9, 2026, under analysis and experiment plans committed before any of them ran (\texttt{docs/REVISION\_R1.md} and \texttt{docs/REVISION\_EXPERIMENTS.md} in the repository).

Replicates of one task are not independent observations. The intraclass correlation of the paired differences within a task reaches 0.95 (a design effect of 12.4) for qwen3-coder's correctness and 0.78 (10.3) for Sonnet's output tokens. Sonnet's 468 token pairs thus carry about as much information as 45 independent ones. Inference about tasks in general therefore rests on the 36 tasks rather than on the replicates. In statistical terms, the target population is the set of tasks these templates can generate: model-building base-R scripts of these six sizes with these kinds of injected flaw. The 36 tasks are the sample drawn from it, and each replicate is a repeated measurement of one sampled task. The inferences generalize from the sample to that population, as the items of a language study generalize to theirs \cite{clark_language_1973}. A pilot calculation sized the study before the full runs, but it sized pairs as if they were independent, so it does not apply to this analysis. Nothing sized the 36 tasks, which the task bank fixed. Appendix~\ref{app:power} gives that calculation and the design's sensitivity, which is known only after the fact.

The unit of inference is the task, and the analysis has two stages \cite{matthews_serial_1990}. In the first stage, the replicates of each model, task, and arm are averaged into the share of correct runs and the mean log value of each token, cost, and time outcome. In the second stage, the arms are compared across tasks, so each task contributes one paired difference and no replicate counts as an independent observation. This task-level analysis replaces the run-level tests of our first analysis. Every test in this paper is a task-level test of one of three null hypotheses, stated once here and referred to by number below, and each Results subsection names the parameters of its own family.

\emph{The arm contrast.} For one model $m$, one outcome $y$, and two arms $A$ and $B$, every comparison tests
\begin{equation}\label{eq:arm}
\begin{aligned}
  H_0 &: \theta_{m,y,AB} = 0 \\
  H_1 &: \theta_{m,y,AB} \neq 0
\end{aligned}
\end{equation}
where $\theta_{m,y,AB}$ is the mean, over the population of tasks, of the task-level difference between arm $A$ and arm $B$ in outcome $y$ for model $m$ (Appendix~\ref{app:twostage}). When we have evidence to reject $H_0$, the two arms differ in outcome $y$ for model $m$, averaged over the population of tasks, in the direction of the sign of $\theta_{m,y,AB}$. This family covers the main experiment (treatment against control, 36 tasks), the function-patch wave (Appendix~\ref{app:patch}), the decomposition and misplaced-anchor experiments (each new arm against its reference, 18 tasks), and the Claude new arms (exploratory, Appendix~\ref{app:claude_r1}). We test $H_0$ with a sign-flip permutation test \cite{good_permutation_2005}, exact over all sign patterns for up to 20 tasks and with 100{,}000 random patterns for 36. The Wilcoxon signed-rank test \cite{wilcoxon_individual_1945} repeats each test with ranks, as a sensitivity analysis \cite{thabane_sensitivity_2013}. Holm's procedure \cite{holm_simple_1979} adjusts the $p$-values within each family of related tests, and we have evidence to reject $H_0$ when the adjusted value $p_H$ is below $\alpha = 0.05$. A correctness conclusion needs this evidence under both scorings, and only effects with evidence enter the abstract and the conclusion. Intervals are 95\% percentile intervals from a cluster bootstrap that resamples tasks within tiers \cite{field_bootstrapping_2007}.

\emph{The aggregate change.} For a token, cost, or time outcome $y$, model $m$, and arms $A$ and $B$, we also test
\begin{equation}\label{eq:agg}
\begin{aligned}
  H_0 &: \gamma_{m,y,AB} = 1 \\
  H_1 &: \gamma_{m,y,AB} \neq 1
\end{aligned}
\end{equation}
where $\gamma_{m,y,AB}$ is the ratio of arm $A$'s total of outcome $y$ over the tasks to arm $B$'s total. The total is the right summary when it has meaning in itself \cite{smith_characterizing_1988}, such as the tokens or the bill of a whole workload. The test, added during the revision, is a permutation test that swaps the two arms within tasks (Appendix~\ref{app:twostage}). When we have evidence to reject $H_0$, the arms' totals over the tasks differ, which is a statement about the workload as a whole rather than about the typical task.

For token, cost, and time outcomes we report two estimates of each effect side by side, because they answer different questions and neither is best for both. The \emph{typical-task} change, $\exp(\hat{\theta}) - 1$, back-transforms the estimate of $\theta$ in Hypotheses~\eqref{eq:arm}, the mean of the task-level log differences. It is the geometric mean of the per-task ratios, the consistent way to average ratios over a set of benchmarks \cite{fleming_how_1986}, and the effect a log-scale analysis estimates when effects are multiplicative \cite{keene_log_1995}. It weighs every task equally, as an inference about the population of tasks requires, so a few long tasks cannot dominate it, and it is the quantity the pre-registered test addresses, which makes it primary. The \emph{aggregate} change, $\hat{\gamma} - 1$, estimates the ratio of totals in Hypotheses~\eqref{eq:agg} and weights long tasks more. The two can differ, even in sign, when the effect reverses between small and large files.

\emph{The tier effect.} For the data pooled over models or for one model, we test
\begin{equation}\label{eq:tier}
\begin{aligned}
  H_0 &: \theta_{\mathrm{T1}} = \theta_{\mathrm{T2}} = \cdots = \theta_{\mathrm{T6}} \\
  H_1 &: \theta_{\mathrm{T}j} \neq \theta_{\mathrm{T}k} \text{ for at least one pair of tiers } j, k
\end{aligned}
\end{equation}
where $\theta_{\mathrm{T}j}$ is the arm contrast of Hypotheses~\eqref{eq:arm} among the tasks of tier T$j$. The test is a permutation test of the tier labels (Sections~\ref{sec:interaction} and~\ref{sec:contractctl}, Appendix~\ref{app:splitplot}). When we have evidence to reject $H_0$, the arm contrast differs between at least two tiers, that is, the effect depends on file size. Tiers group tasks but never replace them as the unit: a tier-group result compares the arms across its 18 tasks, and per-tier values describe six tasks each without a test (Section~\ref{sec:safeguards}). Descriptive comparisons, such as the component arms of the decomposition experiment, carry nominal $p$-values or none.

The permutation test of Hypotheses~\eqref{eq:tier} reassigns the tier labels of the 36 task-level differences 100{,}000 times, and the $p$-value is the share of reassignments whose between-tier variation is at least the observed one. The test uses one value per task and never treats runs as observations, and it makes no distributional assumption. The analysis plan initially called for split-plot analysis of variance (ANOVA) \cite{montgomery_design_2020} on the task means for this question, whose test is the same as a one-way ANOVA of the task-level differences across tiers. Its equal-variance assumption fails in these data, so the permutation test is primary and the $F$ test, Welch's ANOVA \cite{welch_comparison_1951}, and the Kruskal--Wallis test \cite{kruskal_ranks_1952} are reported as nominal checks.

To check that the conclusions do not depend on averaging the replicates first, we also fit hierarchical (mixed-effects) models to the individual runs of each model and contrast \cite{baayen_mixed_2008, bates_lme4_2015}, with a random intercept and arm effect for each task \cite{barr_random_2013}. Token counts are modeled with a negative-binomial count model \cite{hilbe_negative_2011}, with a quasi-Poisson cluster-robust backup where it does not converge \cite{pustejovsky_small_2018, verhoef_quasipoisson_2007}, and correctness with a linear probability model \cite{gomila_logistic_2021}. Their error assumptions fail even on the log scale and under the count model (Appendix~\ref{app:hier}), so their $p$-values are nominal checks at best. They show whether the run-level fits point the same way as the task-level tests, and the conclusions rest on the task-level permutation tests. We added these checks after the results were known (Appendix~\ref{app:stats}).

\subsubsection{Design Safeguards}
\label{sec:safeguards}
The main experiment's comparison rests on five standard components of a well-designed experiment \cite{montgomery_design_2020}.
\begin{itemize}
  \item One manipulated contrast: both arms are rendered from the same change list, so they request the same changes, and they differ in wording and format together (Section~\ref{sec:conditions}).
  \item Blocking through pairing: each task, replicate, and model runs under both arms with the same seed, so task difficulty, model identity, and replicate noise cancel within pairs.
  \item Replication: 36 tasks, six per tier, are the independent units. Each is measured 13 times per arm (the replicates), which lowers within-task noise but adds no independent units (Section~\ref{sec:power}).
  \item Blind, identical measurement: one sandbox and one validator score both arms with no knowledge of condition.
  \item Prespecification and complete reporting: the 6{,}552 runs were planned and their count fixed before the full runs, although the calculation behind that count does not apply to the task-level analysis (Appendix~\ref{app:power}). The revision's analyses were committed before they ran, all planned runs completed, infrastructure failures were re-run with the arms interleaved, and both output contracts are reported.
\end{itemize}
The pipeline, raw outputs, and analysis code are public, so these properties can be verified. These safeguards describe the main experiment. Section~\ref{sec:newarms} states which of them the revision experiments keep.

The design also has limits. The number of independent units, 36 tasks, six per tier, was fixed by the task bank and not by a sample-size calculation (Section~\ref{sec:power}), so the design's sensitivity is reported after the fact (Appendix~\ref{app:power}). Conclusions are drawn over all 36 tasks or over the two tier groups of 18. A single tier offers only six paired differences. Its sign-flip test can reach a two-sided $p$ of 0.031 at best and no Holm-adjusted $p$ below 0.19 across six tiers, and it can detect only effects of about 1.4 standard deviations of the task-level differences (Appendix~\ref{app:power}). Per-tier results are therefore descriptive. Tier is a property of each task rather than a factor assigned at random, and the output contract changes with it (Section~\ref{sec:contract}). 

Two qualifications bound the findings. First, the main treatment is the FileMark-style representation as a whole: wording, line anchor, quoted span, and layout, which together also lengthen the input. The decomposition experiment separates the wording from the format. Second, the findings are scoped to this task population and these models, on this hardware for the local models and on the provider's serving infrastructure during the study window for the hosted ones. Section~\ref{sec:threats} discusses generalization.

\begin{figure*}[!t]
  \centering
  \begin{adjustbox}{max width=\textwidth}
\begin{tikzpicture}[
    font=\small\bfseries,
    >={Stealth[length=2.6mm]},
    node distance=4mm and 5mm,
    box/.style={draw, line width=0.5pt, rounded corners=2pt, align=center,
                inner sep=3pt, minimum height=9mm, fill=gray!5, text=black,
                text width=27mm},
    armbox/.style={box, fill=green!8, text width=24mm, minimum height=7mm},
    arr/.style={->, line width=1pt, black!70}
  ]
  \node[box] (tmpl) {parameterized\\templates\\(tiers T1--T6)};
  \node[box, right=of tmpl, fill=blue!6] (tasks) {36 tasks:\\\texttt{seed.R} $+$\\\texttt{changes.json}};
  \node[armbox, right=6mm of tasks, yshift=6mm] (ctl) {CONTROL\\holistic list};
  \node[armbox, right=6mm of tasks, yshift=-6mm] (trt) {TREATMENT\\FileMark export};
  \node[box, right=6mm of tasks, xshift=31mm, fill=orange!8] (models) {7 models, paired:\\Ollama $\times$5,\\Claude $\times$2};
  \node[box, below=10mm of models, fill=gray!12, text width=44mm] (gen) {output contract,\\same in both arms:\\T1--T3 rewrite,\\T4--T6 edit blocks};
  \node[box, left=of gen, fill=red!6, text width=34mm] (val) {sandboxed run $+$\\graded validation $\Rightarrow$\\correctness, tokens};
  \node[box, left=of val, text width=30mm] (analysis) {task-level\\analysis};
  \draw[arr] (tmpl) -- (tasks);
  \draw[arr] (tasks.east) -- (ctl.west);
  \draw[arr] (tasks.east) -- (trt.west);
  \draw[arr] (ctl.east) -- (models.west);
  \draw[arr] (trt.east) -- (models.west);
  \draw[arr] (models) -- (gen);
  \draw[arr] (gen) -- (val);
  \draw[arr] (val) -- (analysis);
\end{tikzpicture}\end{adjustbox}
  \caption{Experiment pipeline. Both arms are derived from the same \texttt{changes.json}: they request the same changes and differ in wording and format (Section~\ref{sec:conditions}).}
  \label{fig:workflow}
\end{figure*}

\subsection{Revision Experiments}
\label{sec:newarms}
Three experiments, designed and committed before they ran, address what the main contrast cannot separate. They were run during the revision, in October 2026, after the referees' questions were synthesized (Appendix~\ref{app:timeline}). Each runs every model with 3 replicates per arm. The decomposition and misplaced-anchor experiments run on the 18 large tasks (T4 to T6), and each of their arms is compared with the main experiment's control or treatment (13 replicates) on the same tasks. The contract-control experiment runs both of the main experiment's arms under the other output contract on T1 to T4. Table~\ref{tab:arms} lists the arms. Every arm keeps the main experiment's instructions, output contract, and full script, and changes only the block of requested changes.

\begin{table*}[!t]
\caption{Arms of the main experiment and of the revision experiments. Every arm requests the same changes, and only the wording or the format of the requested-changes block, or the output contract, differs.}
\label{tab:arms}
\centering
\footnotesize
\begin{fittable}
\begin{tabular}{>{\raggedright\arraybackslash}p{22mm}>{\raggedright\arraybackslash}p{26mm}>{\raggedright\arraybackslash}p{19mm}>{\raggedright\arraybackslash}p{38mm}>{\raggedright\arraybackslash}p{26mm}}
\toprule
Experiment & Arm & Wording & Format & Output contract \\
\midrule
Main experiment & control & goal statement & plain list & full rewrites (T1--T3), edit blocks (T4--T6) \\
 & treatment & diagnosis and fix & FileMark export (anchor and quoted line) & as above \\
\addlinespace
Contract control & control & goal statement & plain list & edit blocks (T1--T3), full rewrites (T4) \\
 & treatment & diagnosis and fix & FileMark export (anchor and quoted line) & as above \\
\addlinespace
Decomposition & feedback as list & diagnosis and fix & plain list & edit blocks (T4--T6) \\
 & control text, export & goal statement & FileMark export & edit blocks (T4--T6) \\
 & list + line numbers & diagnosis and fix & plain list with each item's line number & edit blocks (T4--T6) \\
 & list + quoted line & diagnosis and fix & plain list with each item's quoted line & edit blocks (T4--T6) \\
\addlinespace
Misplaced anchors & anchor 5 lines off & diagnosis and fix & export, anchor and quote moved 5 lines & edit blocks (T4--T6) \\
 & anchor on the caller line & diagnosis and fix & export, anchor on the driver line that calls the function & edit blocks (T4--T6) \\
 & anchor on the return line & diagnosis and fix & export, anchor on the flawed function's return line & edit blocks (T4--T6) \\
\bottomrule
\end{tabular}
\end{fittable}
\end{table*}

The \emph{decomposition} experiment crosses wording with format. The \emph{misplaced-anchor} experiment keeps the treatment's wording and moves each comment's anchor, with its quoted line, to one of three wrong places. Every library script consists of numbered functions, \texttt{analyze\_block\_1} to \texttt{analyze\_block\_K}, each of which computes one analysis block and ends with a line that returns its results as a list. A \emph{driver} at the end of the file builds the data and calls each function once. The \emph{caller line} is the driver line that calls the function containing the flaw. It names the right function but lies outside it, the anchor on the caller that the referee asked for. The \emph{return line} is the last line of the function containing the flaw. It is where a wrong result leaves the function, the place a reviewer who saw only the symptom might point. The \emph{offset anchor} lies five lines below the flawed line, skipping blank and brace-only lines and never landing on a flawed or a return line, and 77 of its 78 anchors land in a neighboring function. Table~\ref{tab:anchors} shows the four positions for one comment, which asks for group means instead of medians, and the comment's text is the same in every arm. The \emph{contract-control} experiment runs the main experiment's arms under the edit-block contract on T1 to T3, so that all six tiers share one contract, and under full rewrites on T4, so that both contracts meet at one size.

\begin{table*}[!t]
\caption{Misplaced-anchor experiment: where each arm anchors comment 1 of task \texttt{t4\_lib\_01}, by seed line, each anchor in the color its arm keeps in Fig.~\ref{fig:misplaced} \cite{carr_micromaps_2010}. The comment's text is the same in every arm, and only its line number and the quoted line move.}
\label{tab:anchors}
\centering
\begin{fittable}
\begin{tabular}{rll}
\toprule
Line & Script (excerpt of \texttt{seed.R}) & Anchor of the arm \\
\midrule
34 & \texttt{analyze\_block\_3~<-{}~function(d)~\{} & \\
35 & \texttt{~~\#~group-{}wise~summary~of~y~by~factor~g} & \\
\rowcolor{anchoracc!15} 36 & \texttt{~~gm~<-{}~tapply(d\$y,~d\$g,~median)~~~~~~~\#~[block~3]} & \textcolor{anchoracc}{\textbf{accurate anchor (main treatment)}} \\
37 & \texttt{~~counts~<-{}~table(d\$g)} & \\
\rowcolor{anchorreturn!15} 38 & \texttt{~~list(gm~=~gm,~counts~=~counts)} & \textcolor{anchorreturn}{\textbf{return line}} \\
39 & \texttt{\}} & \\
40 &  & \\
\rowcolor{anchoroff!15} 41 & \texttt{analyze\_block\_4~<-{}~function(d)~\{} & \textcolor{anchoroff}{\textbf{anchor 5 lines off}} \\
\addlinespace
162 & \texttt{\#~-{}-{}-{}-{}~driver~-{}-{}-{}-{}} & \\
163 & \texttt{set.seed(42)} & \\
164 & \texttt{d~<-{}~make\_data()} & \\
165 & \texttt{res\_1~<-{}~analyze\_block\_1(d)} & \\
166 & \texttt{res\_2~<-{}~analyze\_block\_2(d)} & \\
\rowcolor{anchorcaller!15} 167 & \texttt{res\_3~<-{}~analyze\_block\_3(d)} & \textcolor{anchorcaller}{\textbf{caller line}} \\
\bottomrule
\end{tabular}
\end{fittable}
\end{table*}

Every contrast of the first two experiments is a parameter of the family of Section~\ref{sec:power}. For model $m$ and outcome $y$, $\theta_{m,y,k}$ is the mean over the 18 large tasks of the task-level difference, first arm minus second arm, for the contrast $k$ of Table~\ref{tab:contrasts}. The outcomes are the share of correct runs, under both scorings, and the output tokens. For the contract-control experiment the question is the file-size question of Section~\ref{sec:interaction} under one contract.

\begin{table*}[!t]
\caption{Decomposition and misplaced-anchor experiments: the contrasts tested, first arm minus second, and the feature of the main treatment each isolates. The feature is in the first arm for F, W, Q, and L, and in the second arm, the treatment with its accurate anchor, for the misplaced anchors.}
\label{tab:contrasts}
\centering
\begin{fittable}
\begin{tabular}{lllll}
\toprule
Contrast & First arm & Second arm & Isolates & Scope \\
\midrule
F & treatment (FileMark export) & feedback as list & the format, at fixed wording & confirmatory \\
W & feedback as list & control & the wording, at fixed format & confirmatory \\
Q & list + quoted line & feedback as list & the quoted line & descriptive \\
L & list + line numbers & feedback as list & the line number & descriptive \\
\addlinespace
off5 & anchor 5 lines off & treatment & an accurate anchor & confirmatory \\
caller & anchor on the caller line & treatment & an accurate anchor & confirmatory \\
return & anchor on the return line & treatment & an accurate anchor & confirmatory \\
\bottomrule
\end{tabular}
\end{fittable}
\end{table*}

The five local models are the confirmatory scope. Their references come from the same period as the new arms, because the main experiment's damaged local cells were re-run alongside them. Holm correction runs within each experiment and outcome across those five models (Appendix~\ref{app:holm}), and correctness findings must hold under both scorings. The two Claude models ran the same arms against their July references, three months older, so their results are exploratory and reported in Appendix~\ref{app:claude_r1}. The smallest effects the new arms can detect are given in Appendix~\ref{app:power}.

Of the main experiment's safeguards (Section~\ref{sec:safeguards}), the revision experiments keep the manipulated contrast, which again changes only the block of requested changes, and the blind and identical measurement. They keep the prespecification, since their designs and analyses were committed before they ran. They keep the pairing only at the task level: a new arm's 3 replicates are compared with the reference arm's 13 on the same tasks, so replicate noise does not cancel within pairs. The independent units are the 18 large tasks rather than 36, each measured 3 rather than 13 times per arm, and for the Claude models the reference arms were run three months earlier, which is why their results are exploratory.

\section{Results}
\label{sec:results}
This section reports the main experiment (Section~\ref{sec:results_main}) and then the revision experiments (Section~\ref{sec:newresults}). All results use the corrected data, in which the main experiment's 2{,}652 local runs from the cells damaged by the context truncation are replaced by their October re-runs (Section~\ref{sec:env}, Appendix~\ref{app:timeline}). Every effect is a task-level comparison of two arms, estimated as in Section~\ref{sec:power}. In the tables, bold with an asterisk marks evidence to reject the null hypothesis of no difference, and an underline marks evidence in favor of the main treatment or of the arm with the feature under test (Table~\ref{tab:contrasts}). 

\subsection{Main Experiment}
\label{sec:results_main}

\subsubsection{Tokens and Cost}
\label{sec:tokens}
For each model $m$ and outcome $y$ we test the arm contrast in Hypotheses~\eqref{eq:arm}, where $\theta_{m,y}$ denotes $\theta_{m,y,AB}$ with the treatment as $A$ and the control as $B$. The outcome $y$ is the total tokens, the output tokens, the cost of a call, or the seconds per call, over all 36 tasks (Table~\ref{tab:main}) or within a tier group (Table~\ref{tab:cost}). For the token, cost, and time outcomes we also test the aggregate change in Hypotheses~\eqref{eq:agg}.

The export lengthens the request and shortens the answer. The task prompt grows by 28.6\% on T1 to T3 and by 4.4\% on T4 to T6, counted with one tokenizer. 
Over all tasks the net effect on total tokens is small (Table~\ref{tab:main}). For the typical task, four models have evidence to reject $H_0$, all for increases of 6.9\% to 14.7\% (Holm $p \le 0.047$), and the other three do not. File size explains the pattern (Table~\ref{tab:cost}, Fig.~\ref{fig:tier}a), which is why we report the two tier groups, split where the output contract changes (Section~\ref{sec:contract}). On T1 to T3, where the longer request is a large share of a short exchange, total tokens rose for every model, with evidence to reject $H_0$ for six of the seven. On T4 to T6 they fell for four models, with evidence to reject $H_0$ for both Claude models and gemma4 (by up to 18.3\%, for Sonnet), and rose by 4.1\% to 8.4\% for the other three. The aggregate change, which weights the long tasks more, is lower than the typical-task change for every model. 
Claude Code's own context, 12{,}000 to 49{,}000 input tokens per call in both arms, keeps the hosted totals as run within 3\% of control in both tier groups. 

\begin{figure*}[!t]
  \centering
  \includegraphics[width=0.8\textwidth]{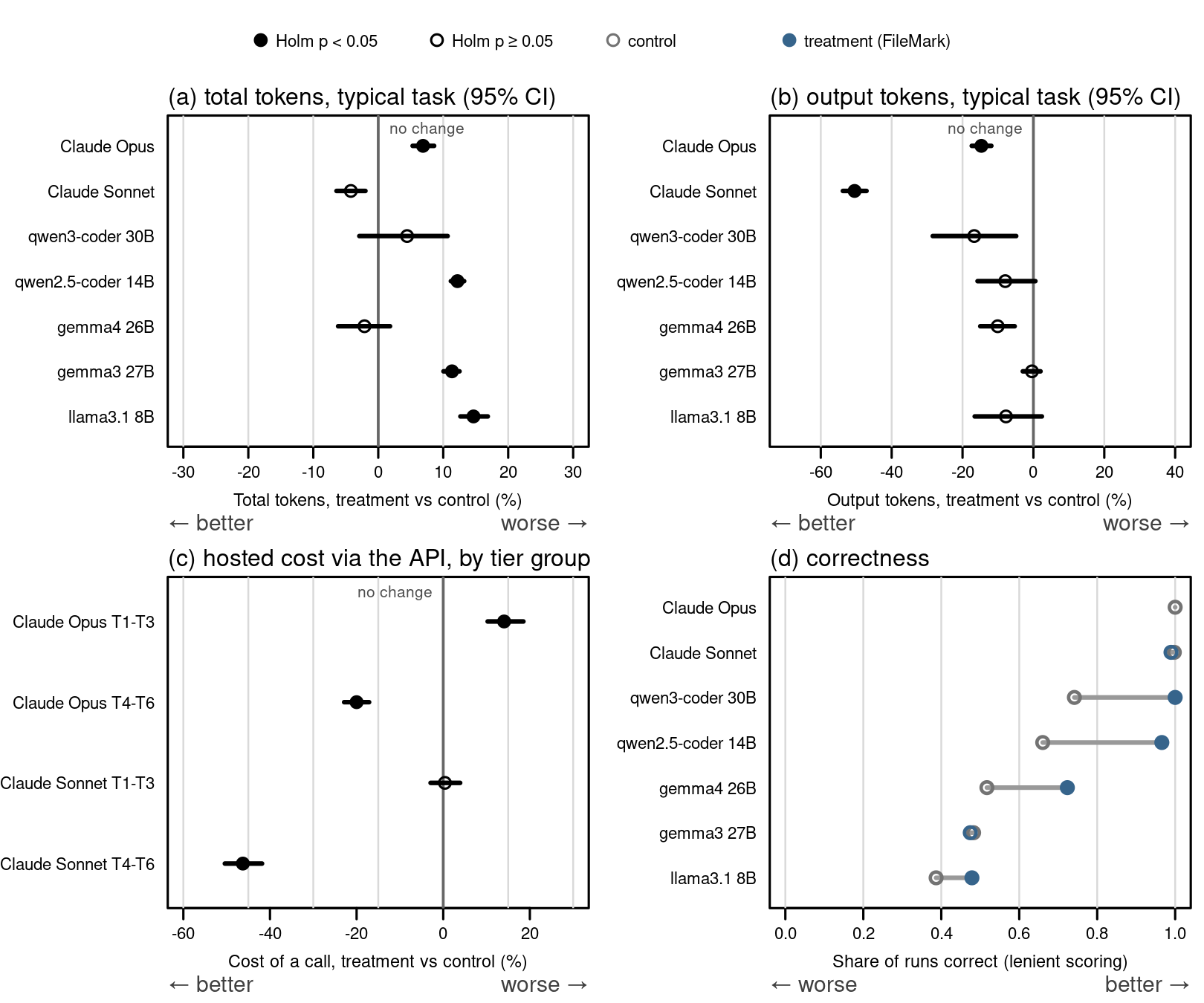}
  \caption{Main experiment, task level: the treatment changed the typical task's total tokens little (a), shortened the Claude models' answers (b), lowered the hosted cost of an edit on T4--T6 and raised it on T1--T3 (c), and raised four local models' correctness (d). A filled point marks evidence to reject the null hypothesis of no effect (Holm $p < 0.05$).}
  \label{fig:main}
\end{figure*}

The answer is where the treatment saves (Fig.~\ref{fig:main}b). Both Claude models have evidence to reject $H_0$ for output tokens (Holm $p \le 0.014$): the typical answer shrank by 14.7\% (Opus) and 50.5\% (Sonnet), almost all of it on T4 to T6, where it shrank by 34.1\% and 74.1\%. We do not have evidence to reject $H_0$ for the typical local model answer.  However, three answers shrank in aggregate, where the long answers on large files weigh most. 

Cost follows the output tokens, because an output token costs five times as much as an input token (Table~\ref{tab:costparts}). Via the API, the typical T4 to T6 edit cost 20.0\% (Opus) and 46.3\% (Sonnet) less, and both models have evidence to reject $H_0$ (Holm $p \le 2 \times 10^{-5}$). On T1 to T3 it cost 14.1\% more for Opus, for which we have evidence to reject $H_0$, and about the same for Sonnet. As run, the interface's own context dilutes these changes to $-5.0\%$ and $-10.5\%$ on T4 to T6. The time per call followed the answer length, falling by 26.7\% (Opus) and 61.7\% (Sonnet) for the typical T4 to T6 call. Energy was not measured, so an energy saving is an implication of the shorter answers, not a result. 

\begin{table*}[!t]
\caption{Main experiment, tokens, task level (36 tasks): treatment minus control for the typical task (sign-flip test) and in aggregate (permutation test of the ratio of totals), with 95\% cluster-bootstrap intervals and Holm-adjusted $p$-values ($p_H$). Bold with an asterisk marks evidence to reject the null hypothesis of no difference, underlined where the treatment used fewer tokens.}
\label{tab:main}
\centering
\setlength{\tabcolsep}{2pt}
\begin{fittable}
\begin{tabular}{lrrrrrrrr}
\toprule
 & \multicolumn{4}{c}{Output tokens, change (\%)} & \multicolumn{4}{c}{Total tokens, change (\%)} \\
\cmidrule(lr){2-5}\cmidrule(lr){6-9}
 & \multicolumn{2}{c}{Typical task} & \multicolumn{2}{c}{Aggregate} & \multicolumn{2}{c}{Typical task} & \multicolumn{2}{c}{Aggregate} \\
\cmidrule(lr){2-3}\cmidrule(lr){4-5}\cmidrule(lr){6-7}\cmidrule(lr){8-9}
Model & \hd{Change}{(95\% CI)} & $p_H$ & \hd{Change}{(95\% CI)} & $p_H$ & \hd{Change}{(95\% CI)} & $p_H$ & \hd{Change}{(95\% CI)} & $p_H$ \\
\midrule
Claude Opus & \sigci{$+6.9$}{$+5.3$}{$+8.6$} & \sigp{0.047} & \ci{$-1.7$}{$-2.9$}{$-0.5$} & 0.44 & \goodci{$-14.7$}{$-17.4$}{$-11.9$} & \sigp{0.014} & \goodci{$-22.5$}{$-25.6$}{$-19.3$} & \sigp{0.0006} \\
Claude Sonnet & \ci{$-4.2$}{$-6.4$}{$-2.0$} & 0.62 & \goodci{$-11.7$}{$-14.3$}{$-9.1$} & \sigp{0.0010} & \goodci{$-50.5$}{$-53.8$}{$-47.1$} & \sigp{$<10^{-4}$} & \goodci{$-57.5$}{$-61.3$}{$-53.0$} & \sigp{$<10^{-4}$} \\
\addlinespace
qwen3-coder 30B & \ci{$+4.4$}{$-2.9$}{$+10.7$} & 0.78 & \ci{$-5.2$}{$-13.8$}{$+3.1$} & 0.46 & \ci{$-16.7$}{$-28.4$}{$-4.9$} & 0.13 & \goodci{$-44.0$}{$-61.0$}{$-15.1$} & \sigp{0.03} \\
qwen2.5-coder 14B & \sigci{$+12.2$}{$+11.2$}{$+13.2$} & \sigp{$<10^{-4}$} & \sigci{$+4.9$}{$+4.2$}{$+5.6$} & \sigp{$<10^{-4}$} & \ci{$-7.9$}{$-15.8$}{$+0.5$} & 0.28 & \goodci{$-10.0$}{$-16.1$}{$-3.7$} & \sigp{0.038} \\
gemma4 26B & \ci{$-2.1$}{$-6.2$}{$+1.8$} & 0.78 & \goodci{$-5.7$}{$-7.5$}{$-3.9$} & \sigp{0.0033} & \ci{$-10.1$}{$-15.0$}{$-5.3$} & 0.053 & \goodci{$-12.9$}{$-15.8$}{$-10.1$} & \sigp{0.0002} \\
llama3.1 8B & \sigci{$+14.7$}{$+12.6$}{$+16.9$} & \sigp{$<10^{-4}$} & \sigci{$+5.9$}{$+2.9$}{$+9.3$} & \sigp{0.018} & \ci{$-7.7$}{$-16.6$}{$+2.4$} & 0.73 & \ci{$+1.4$}{$-20.9$}{$+29.5$} & 0.94 \\
gemma3 27B & \sigci{$+11.4$}{$+10.0$}{$+12.5$} & \sigp{$<10^{-4}$} & \sigci{$+4.2$}{$+3.5$}{$+5.1$} & \sigp{$<10^{-4}$} & \ci{$-0.4$}{$-3.0$}{$+2.0$} & 0.81 & \ci{$+1.7$}{$+0.1$}{$+3.4$} & 0.42 \\
\bottomrule
\end{tabular}
\end{fittable}
\end{table*}

\begin{table*}[!t]
\caption{Main experiment by tier group: treatment minus control for the typical task (sign-flip test, 95\% cluster-bootstrap intervals) and in aggregate (permutation test), with Holm-adjusted $p$-values in \texttt{results/token\_estimands/main.csv} and cost for the hosted models only. Bold with an asterisk marks evidence to reject the null hypothesis of no difference, underlined where the treatment used fewer tokens, money, or time.}
\label{tab:cost}
\centering
\footnotesize
\setlength{\tabcolsep}{4pt}
\begin{fittable}
\begin{tabular}{lrr>{\columncolor{tablegray}}r>{\columncolor{tablegray}}r}
\toprule
 & \multicolumn{2}{c}{T1--T3 (18 tasks)} & \multicolumn{2}{c}{T4--T6 (18 tasks)} \\
\cmidrule(lr){2-3}\cmidrule(lr){4-5}
Model & \hd{Typical}{(95\% CI)} & Aggregate & \hd{Typical}{(95\% CI)} & Aggregate \\
\midrule
\multicolumn{5}{l}{\emph{Total tokens, change (\%)}} \\
Claude Opus & \sigcii{$+21.2$}{$+18.3$}{$+24.4$} & \sig{$+17.1$} & \goodcii{$-5.7$}{$-7.4$}{$-4.1$} & \good{$-4.8$} \\ 
Claude Sonnet & \sigcii{$+12.3$}{$+9.4$}{$+15.6$} & \sig{$+9.7$} & \goodcii{$-18.3$}{$-21.3$}{$-15.3$} & \good{$-15.2$} \\ 
\addlinespace
qwen3-coder 30B & \sigcii{$+19.9$}{$+16.9$}{$+22.5$} & \sig{$+15.8$} & \cii{$-9.0$}{$-21.2$}{$+2.1$} & $-8.0$ \\ 
qwen2.5-coder 14B & \sigcii{$+20.4$}{$+18.7$}{$+22.3$} & \sig{$+16.4$} & \sigcii{$+4.5$}{$+3.3$}{$+5.5$} & \sig{$+3.1$} \\ 
gemma4 26B & \cii{$+5.0$}{$-3.5$}{$+13.4$} & $-0.6$ & \goodcii{$-8.8$}{$-10.4$}{$-7.3$} & \good{$-6.9$} \\ 
llama3.1 8B & \sigcii{$+21.3$}{$+20.0$}{$+22.6$} & \sig{$+17.4$} & \sigcii{$+8.4$}{$+4.7$}{$+12.5$} & $+4.2$ \\ 
gemma3 27B & \sigcii{$+19.1$}{$+16.5$}{$+21.4$} & \sig{$+15.0$} & \sigcii{$+4.1$}{$+3.2$}{$+5.0$} & \sig{$+3.3$} \\ 
\midrule
\multicolumn{5}{l}{\emph{Output tokens, change (\%)}} \\
Claude Opus & \sigcii{$+10.4$}{$+6.5$}{$+15.0$} & \sig{$+8.0$} & \goodcii{$-34.1$}{$-37.4$}{$-30.7$} & \good{$-34.0$} \\ 
Claude Sonnet & \goodcii{$-5.4$}{$-8.6$}{$-1.9$} & \good{$-5.2$} & \goodcii{$-74.1$}{$-77.3$}{$-70.7$} & \good{$-77.4$} \\ 
\addlinespace
qwen3-coder 30B & \cii{$-2.3$}{$-8.0$}{$+2.3$} & $-1.5$ & \cii{$-28.9$}{$-47.7$}{$-7.5$} & \good{$-55.4$} \\ 
qwen2.5-coder 14B & \cii{$+0.5$}{$-2.6$}{$+4.2$} & 0.0 & \cii{$-15.7$}{$-29.2$}{0.0} & \good{$-25.0$} \\ 
gemma4 26B & \cii{$-0.7$}{$-10.6$}{$+9.4$} & $-6.4$ & \goodcii{$-18.6$}{$-21.6$}{$-16.0$} & \good{$-15.3$} \\ 
llama3.1 8B & \cii{$+0.6$}{$-2.5$}{$+4.0$} & $+1.6$ & \cii{$-15.4$}{$-30.7$}{$+3.4$} & $+1.2$ \\ 
gemma3 27B & \cii{$-2.6$}{$-7.4$}{$+1.8$} & $-1.4$ & \cii{$+1.8$}{$+0.2$}{$+3.6$} & $+2.0$ \\ 
\midrule
\multicolumn{5}{l}{\emph{Cost via API, change (\%)}} \\
Claude Opus & \sigcii{$+14.1$}{$+10.3$}{$+18.5$} & \sig{$+10.8$} & \goodcii{$-20.0$}{$-22.9$}{$-17.1$} & \good{$-19.1$} \\ 
Claude Sonnet & \cii{$+0.4$}{$-2.9$}{$+3.9$} & $-0.6$ & \goodcii{$-46.3$}{$-50.4$}{$-41.8$} & \good{$-45.6$} \\ 
\midrule
\multicolumn{5}{l}{\emph{Cost as run, change (\%)}} \\
Claude Opus & \sigcii{$+5.2$}{$+4.6$}{$+5.9$} & \sig{$+5.1$} & \goodcii{$-5.0$}{$-6.0$}{$-3.9$} & \good{$-4.9$} \\ 
Claude Sonnet & \sigcii{$+1.8$}{$+0.9$}{$+2.8$} & \sig{$+1.8$} & \goodcii{$-10.5$}{$-12.4$}{$-8.7$} & \good{$-10.4$} \\ 
\midrule
\multicolumn{5}{l}{\emph{Seconds per call, change (\%)}} \\
Claude Opus & \sigcii{$+9.1$}{$+6.4$}{$+12.2$} & \sig{$+7.8$} & \goodcii{$-26.7$}{$-29.7$}{$-23.7$} & \good{$-26.9$} \\ 
Claude Sonnet & \cii{$-3.3$}{$-6.2$}{$-0.2$} & $-3.7$ & \goodcii{$-61.7$}{$-65.8$}{$-57.2$} & \good{$-65.2$} \\ 
\addlinespace
qwen3-coder 30B & \cii{$-0.9$}{$-6.2$}{$+3.8$} & $-0.4$ & \cii{$-28.8$}{$-47.2$}{$-7.8$} & \good{$-54.8$} \\ 
qwen2.5-coder 14B & \cii{$+1.6$}{$-1.6$}{$+5.4$} & $+1.1$ & \cii{$-14.2$}{$-27.5$}{$+0.7$} & \good{$-24.3$} \\ 
gemma4 26B & \cii{$-0.3$}{$-10.2$}{$+9.9$} & $-6.2$ & \goodcii{$-17.9$}{$-21.0$}{$-15.3$} & \good{$-14.8$} \\ 
llama3.1 8B & \cii{$+1.5$}{$-1.4$}{$+4.8$} & $+2.4$ & \cii{$-13.2$}{$-27.5$}{$+4.7$} & $-1.8$ \\ 
gemma3 27B & \cii{$-1.3$}{$-5.8$}{$+2.7$} & $-0.3$ & \cii{$+2.0$}{$+0.3$}{$+3.7$} & $+2.1$ \\ 
\bottomrule
\end{tabular}
\end{fittable}
\end{table*}

\subsubsection{Correctness}
\label{sec:correctness}
For each model $m$ and each scoring we test Hypotheses~\eqref{eq:arm},
where $\theta^{\mathrm{cor}}_m$ is $\theta_{m,y,AB}$ for treatment against control in model $m$'s share of correct runs, and the pooled test averages that share over the seven models.

\begin{figure*}[!t]
  \centering
  \includegraphics[width=0.8\textwidth]{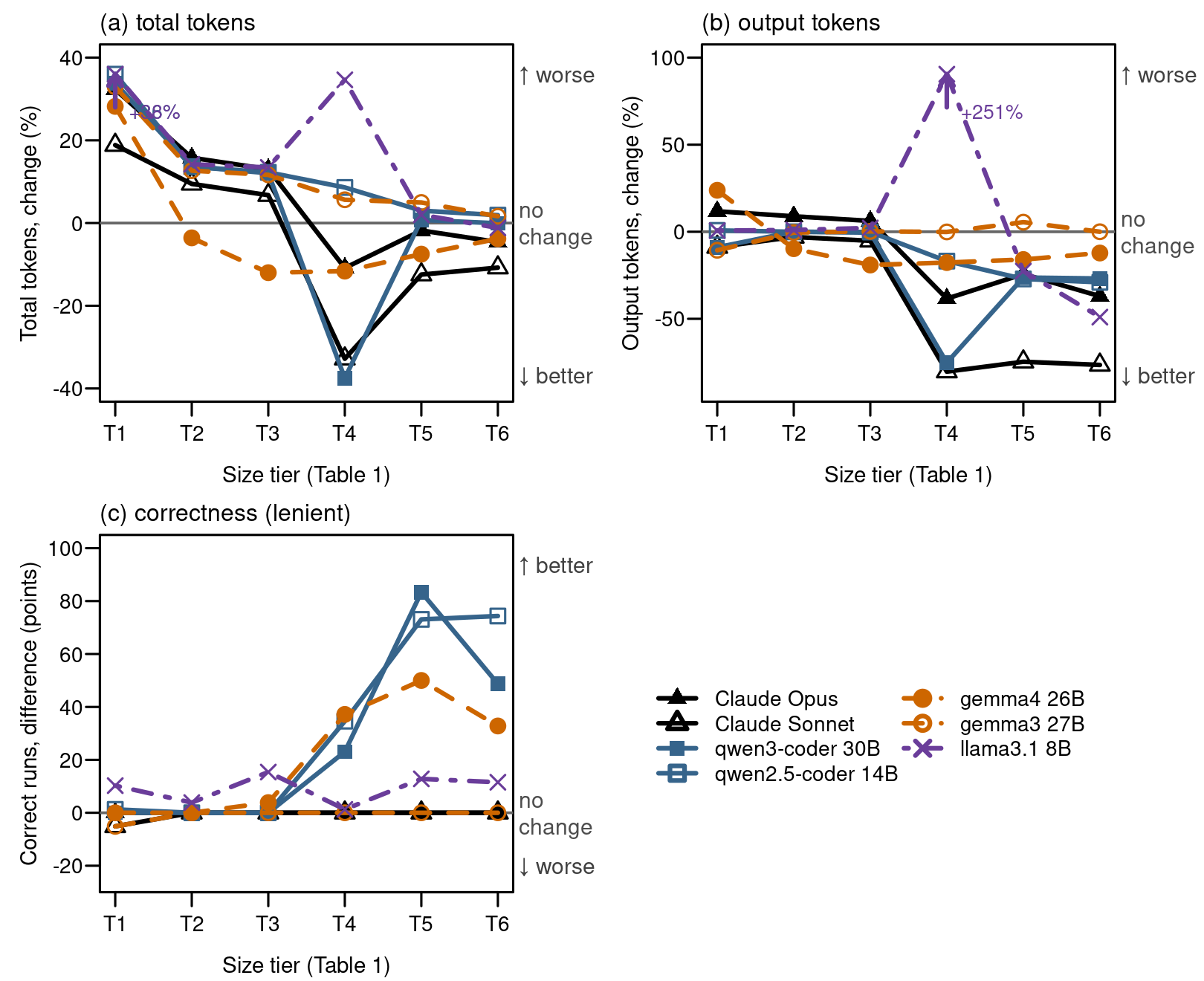}
  \caption{Main experiment by size tier: each point is the aggregate change over the six tasks of a tier, descriptive, with no test per tier (Section~\ref{sec:safeguards}). Total tokens rise on small files and fall on large ones (a), the Claude models' answers shrink on T4--T6 (b), and the coder models and gemma4 gain correctness there (c); off-scale values are capped.}
  \label{fig:tier}
\end{figure*}

Four of the five local models have evidence to reject $H_0$, with identical results under both scorings (Holm $p \le 0.0062$, Fig.~\ref{fig:main}d, Table~\ref{tab:correct}). Their gains were 9.2 to 30.6 points, or 24\% to 46\% of their control rates. 
gemma3 does not have evidence to reject $H_0$. It never produced an edit block on the large files in either arm (Section~\ref{sec:failure}), and Opus was correct in every run of both arms, so its correctness has no test. 
Pooled over models, correctness rose by 11.6 points under strict and 12.1 points under lenient scoring, and by 22.0 and 23.0 points on T4 to T6. In each case we have evidence to reject $H_0$ for the pooled share (sign-flip $p \le 3 \times 10^{-5}$). 
The gains concentrate on the large files, by up to 83 points on T5 (Fig.~\ref{fig:tier}c, descriptive), and the nominal hierarchical checks point the same way for every correctness contrast of the main experiment (Appendix~\ref{app:hier}). 

\begin{table*}[!t]
\caption{Main experiment, correctness, task level (36 tasks): treatment minus control in the share of correct runs, in percentage points, with 95\% cluster-bootstrap intervals and Holm-adjusted sign-flip $p$-values ($p_H$). Bold with an asterisk marks evidence to reject the null hypothesis of no difference, underlined where the treatment was correct more often.}
\label{tab:correct}
\centering
\begin{fittable}
\begin{tabular}{lrrrr}
\toprule
 & \multicolumn{2}{c}{Correct runs, strict} & \multicolumn{2}{c}{Correct runs, lenient} \\
\cmidrule(lr){2-3}\cmidrule(lr){4-5}
Model & \hd{Points}{(95\% CI)} & $p_H$ & \hd{Points}{(95\% CI)} & $p_H$ \\
\midrule
Claude Opus & \ci{0.0}{0.0}{0.0} & n/a & \ci{0.0}{0.0}{0.0} & n/a \\
Claude Sonnet & \sigci{$-4.5$}{$-7.9$}{$-1.5$} & \sigp{0.036} & \ci{$-0.9$}{$-2.6$}{0.0} & 1 \\
\addlinespace
qwen3-coder 30B & \goodci{$+25.9$}{$+16.5$}{$+35.5$} & \sigp{0.0062} & \goodci{$+25.9$}{$+16.5$}{$+35.5$} & \sigp{0.0062} \\
qwen2.5-coder 14B & \goodci{$+30.6$}{$+23.1$}{$+37.8$} & \sigp{0.0002} & \goodci{$+30.6$}{$+23.1$}{$+37.8$} & \sigp{0.0002} \\
gemma4 26B & \goodci{$+20.6$}{$+13.9$}{$+27.5$} & \sigp{0.0004} & \goodci{$+20.6$}{$+13.9$}{$+27.5$} & \sigp{0.0004} \\
llama3.1 8B & \goodci{$+9.2$}{$+5.3$}{$+13.7$} & \sigp{0.0002} & \goodci{$+9.2$}{$+5.3$}{$+13.7$} & \sigp{0.0002} \\
gemma3 27B & \ci{$-0.9$}{$-2.6$}{0.0} & 1 & \ci{$-0.9$}{$-2.6$}{0.0} & 1 \\
\bottomrule
\end{tabular}
\end{fittable}
\end{table*}

Claude Sonnet is the one model whose correctness appeared to fall. Under strict scoring we have evidence to reject $H_0$ (from 0.994 to 0.949, Holm $p = 0.036$). Under lenient scoring we do not have evidence to reject $H_0$ (0.998 to 0.989, Holm $p = 1$), so we do not conclude that the treatment lowered it. 
Of the 27 task--replicate pairs in which exactly one arm failed, 21 (19 treatment, 2 control) were correct edits that the strict scorer missed, because each edit sat in a code block of its own. The treatment invites that form by presenting each change as a separate comment. The other 6 were one misread phrase on one small task (\texttt{t1\_diag\_01}). None of the 27 shows the model losing sight of the wider script: no edit fell outside the requested lines, no requested change was omitted, and no script changed a block it was not asked to change. 

\subsubsection{File Size}
\label{sec:interaction}
The treatment effect depends on file size. For the data pooled over models and for each model, we test the tier effect in Hypotheses~\eqref{eq:tier},
where $\theta_{\mathrm{T}j}$ is the mean task-level difference between treatment and control among the six tasks of tier T$j$. The outcome is the log output tokens or the share of correct runs. The test is the permutation test of Section~\ref{sec:power} on the 36 task-level differences (Appendix~\ref{app:splitplot}).

Pooled over models, we have evidence to reject $H_0$ for output tokens and for correctness under both scorings (permutation $p \le 2 \times 10^{-5}$). Model by model, the token effect depends on tier for both Claude models, gemma4, and llama3.1, and the correctness effect for the two coder models and gemma4, with evidence to reject $H_0$ for each (Holm $p \le 0.0073$). The nominal checks (the pre-registered $F$ test, Welch's ANOVA where it is defined, and the Kruskal--Wallis test) reach the same decisions, and the pooled result generalizes over tasks, not over models. 
Per tier, the picture is descriptive (Fig.~\ref{fig:tier}, Appendix~\ref{app:tiers}): the effects appear at T4, the first tier with the edit-block contract, and the contract-control experiment (Section~\ref{sec:contractctl}) separates the contract from the size.

\subsubsection{Failure Modes and the Edit-Block Contract}
\label{sec:failure}
Table~\ref{tab:failures} classifies every failed local run of the main experiment on T4 to T6. Three models fail mostly on the format, in both arms. gemma3 returned the whole script instead of edit blocks in every run, and on T6 the script also overran the 8{,}192-token output limit. llama3.1 wrote malformed edit markers in 401 of its 468 runs, and gemma4 used its whole output budget on hidden reasoning in 246. 
The treatment changes the failures that depend on finding the code. Across the local models, runs that executed but got the requested changes wrong fell from 199 under control to 18 under treatment. Edit blocks whose \texttt{SEARCH} text matched no line (66, all qwen2.5-coder) and empty answers after long reasoning (178 against 68, gemma4) were also mostly control failures. 
No script in either arm used a package outside base R.

\begin{table*}[!t]
\caption{Main experiment, failed local runs on T4--T6 by cause and arm (lenient scoring, 1{,}170 runs per arm): failures to produce the edit format occur in both arms, and failures to find or fix the code occur mostly under control. The last column names the model with most failed runs in the row (both arms) and its share.}
\label{tab:failures}
\centering
\begin{fittable}
\begin{tabular}{llrrl}
\toprule
Kind & Cause & Control & Treatment & \hd{Model with most failures}{(both arms, share of row)} \\
\midrule
Edit format & Whole script instead of edit blocks & 181 & 161 & gemma3 (91\%) \\
 & Cut off at the output limit & 95 & 90 & gemma3 (84\%) \\
 & Malformed edit markers & 246 & 251 & llama3.1 (81\%) \\
\addlinespace
Finding or fixing code & Reasoning used the whole budget & 178 & 68 & gemma4 (100\%) \\
 & \texttt{SEARCH} text matches no line & 66 & 0 & qwen2.5-coder (100\%) \\
 & Ran, requested changes wrong & 199 & 18 & qwen3-coder (56\%) \\
\addlinespace
Other & No edit block, or runtime error & 2 & 3 & gemma4 (80\%) \\
\midrule
All failures & & 967 & 591 & \\
\bottomrule
\end{tabular}
\end{fittable}
\end{table*}

\subsection{Revision Experiments}
\label{sec:newresults}
The five local models are the confirmatory scope, with Holm correction within each experiment and outcome, and the Claude results are exploratory (Appendix~\ref{app:claude_r1}). Every effect is the first arm minus the second of Table~\ref{tab:contrasts}, and token effects are given for the typical task and, in Table~\ref{tab:newarms}, in aggregate.

\subsubsection{Wording or Format?}
\label{sec:decomp}
\begin{figure*}[!t]
  \centering
  \includegraphics[width=0.8\textwidth]{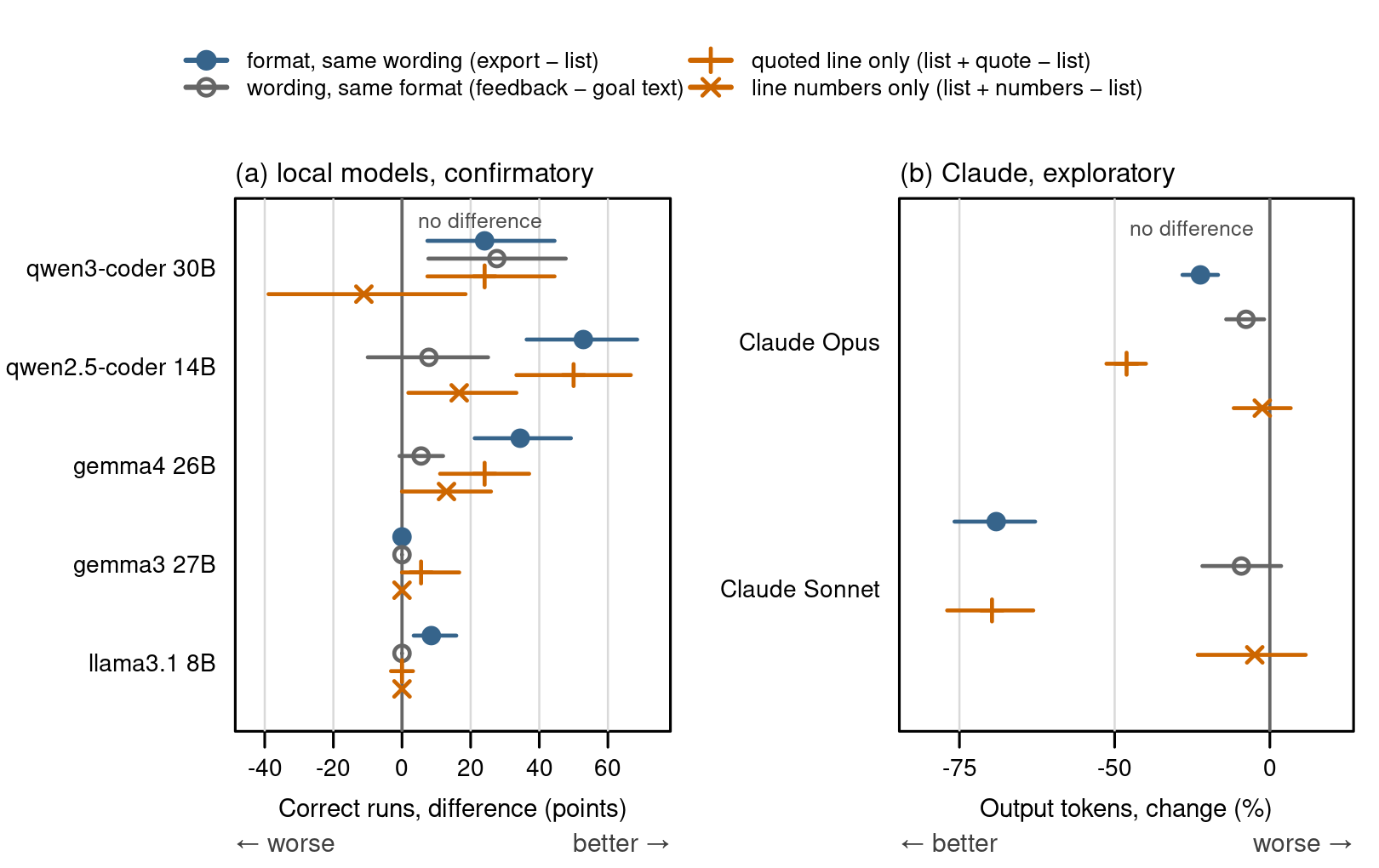}
  \caption{Decomposition experiment (T4--T6), typical-task effects with 95\% intervals: at fixed wording, the FileMark format raised the local models' correctness and cut the Claude models' output tokens, and adding only the quoted line did most of the same.}
  \label{fig:decomp}
\end{figure*}

The treatment changes the wording and the format together (Section~\ref{sec:conditions}). The decomposition experiment separates them on T4 to T6 with four new arms (Fig.~\ref{fig:decomp}, Table~\ref{tab:newarms}). For each local model $m$, outcome $y$, and contrast $k$ of Table~\ref{tab:contrasts} we test Hypotheses~\eqref{eq:arm}, where $\theta_{m,y,k}$ is the mean over the 18 large tasks of the task-level difference, first arm minus second arm. The contrasts are the format at fixed wording (F), the wording at fixed format (W), the quoted line alone (Q), and the line number alone (L).

Holding the wording fixed, the FileMark format raised the correctness of three of the five local models by 8.5 to 52.8 points under both scorings, and these three have evidence to reject $H_0$ for $\theta_{m,y,\mathrm{F}}$ (Holm $p \le 0.02$). The nominal hierarchical checks support the two larger effects but not llama3.1's, which comes from a few tasks (Appendix~\ref{app:hier}). Holding the format fixed, no local model has evidence to reject $H_0$ for $\theta_{m,y,\mathrm{W}}$ (Holm $p \ge 0.12$). The format, not the wording, therefore carries the correctness effect this experiment can detect, although a smaller wording effect cannot be ruled out (Appendix~\ref{app:power}). 

Within the format, the quoted line does most of the work (descriptive, nominal $p$-values). Adding only the quoted line to the plain list raised correctness by 50.0 points for qwen2.5-coder and 24.1 for gemma4, and both have evidence to reject $H_0$ for $\theta_{m,y,\mathrm{Q}}$ at the nominal level ($p \le 0.017$). Adding only the line number changed no model's correctness by more than 17 points, and no model has evidence to reject $H_0$ for $\theta_{m,y,\mathrm{L}}$ ($p \ge 0.094$), as expected for a script shown without line numbers. Exporting the control's own wording in FileMark's format helped three models too, by 29.6 to 51.7 points. 
For the Claude models (exploratory), the format at fixed wording cut the typical task's output tokens by 26.7\% (Opus) and 70.5\% (Sonnet), and the quoted line alone by 47.3\% and 71.1\%. 

\begin{table*}[!t]
\caption{Decomposition experiment (first two rows of each panel) and misplaced-anchor experiment (last three), local models (confirmatory, 18 large tasks): first arm minus second (Table~\ref{tab:contrasts}), with output tokens for the typical task and in aggregate. Bold with an asterisk marks evidence to reject the null hypothesis of no difference, underlined where the arm with the feature under test did better.}
\label{tab:newarms}
\centering
\begin{fittable}
\begin{tabular}{lrrrrr}
\toprule
Contrast (first arm $-$ second arm) & qwen3-coder & qwen2.5-coder & gemma4 & llama3.1 & gemma3 \\
\midrule
\multicolumn{6}{l}{\emph{Correct runs, difference in points (lenient scoring)}} \\
Format: export $-$ list & $+24.1$ & \good{$+52.8$} & \good{$+34.4$} & \good{$+8.5$} & 0 \\ 
Wording: list $-$ control & $+27.6$ & $+7.8$ & $+5.6$ & 0 & 0 \\ 
\addlinespace
5 lines off $-$ accurate & \good{$-68.5$} & \good{$-93.6$} & \good{$-45.5$} & \good{$-8.5$} & $+3.7$ \\ 
Caller line $-$ accurate & 0 & \good{$-54.7$} & \good{$-34.4$} & $-6.7$ & 0 \\ 
Return line $-$ accurate & \good{$-31.5$} & \good{$-71.4$} & \good{$-38.1$} & \good{$-8.5$} & 0 \\ 
\midrule
\multicolumn{6}{l}{\emph{Output tokens, typical-task change in percent}} \\
Format: export $-$ list & $-40.5$ & $-9.8$ & \good{$-17.2$} & $-15.3$ & $+0.7$ \\ 
Wording: list $-$ control & $+19.3$ & $-6.5$ & $-1.6$ & $-0.1$ & $+1.1$ \\ 
\addlinespace
5 lines off $-$ accurate & $+20.1$ & \good{$+204.1$} & \good{$+26.6$} & \good{$+227.6$} & $+0.5$ \\ 
Caller line $-$ accurate & $-1.7$ & $+0.9$ & \good{$+24.0$} & $-16.0$ & $+0.4$ \\ 
Return line $-$ accurate & $+8.2$ & $+12.5$ & \good{$+24.7$} & $+92.6$ & $-0.7$ \\ 
\midrule
\multicolumn{6}{l}{\emph{Output tokens, aggregate change in percent}} \\
Format: export $-$ list & $-64.0$ & $-17.9$ & \good{$-14.0$} & $+9.7$ & $+0.8$ \\ 
Wording: list $-$ control & $+23.8$ & $-8.7$ & $-1.5$ & $-7.8$ & $+1.2$ \\ 
\addlinespace
5 lines off $-$ accurate & $+25.0$ & \good{$+242.2$} & \good{$+21.1$} & \good{$+233.7$} & $+0.1$ \\ 
Caller line $-$ accurate & $+1.1$ & $+8.4$ & \good{$+18.8$} & $-42.8$ & $+0.3$ \\ 
Return line $-$ accurate & $+5.1$ & $+21.1$ & \good{$+19.5$} & $+89.0$ & $-0.8$ \\ 
\bottomrule
\end{tabular}
\end{fittable}
\end{table*}

\subsubsection{Misplaced Anchors}
\label{sec:misplaced}
A misplaced anchor hurts far more than we expected. For each local model $m$, outcome $y$, and misplaced anchor $k$ we test Hypotheses~\eqref{eq:arm},
where $\theta_{m,y,k}$ is the mean over the 18 large tasks of the task-level difference for anchor $k$. The difference is between the arm with anchor $k$ and the treatment arm with the accurate anchor (Table~\ref{tab:contrasts}). The anchor $k$ is five lines off, on the caller line, or on the return line. We had pre-registered small effects, because each comment's text names the exact code.

\begin{figure*}[!t]
  \centering
  \includegraphics[width=0.8\textwidth]{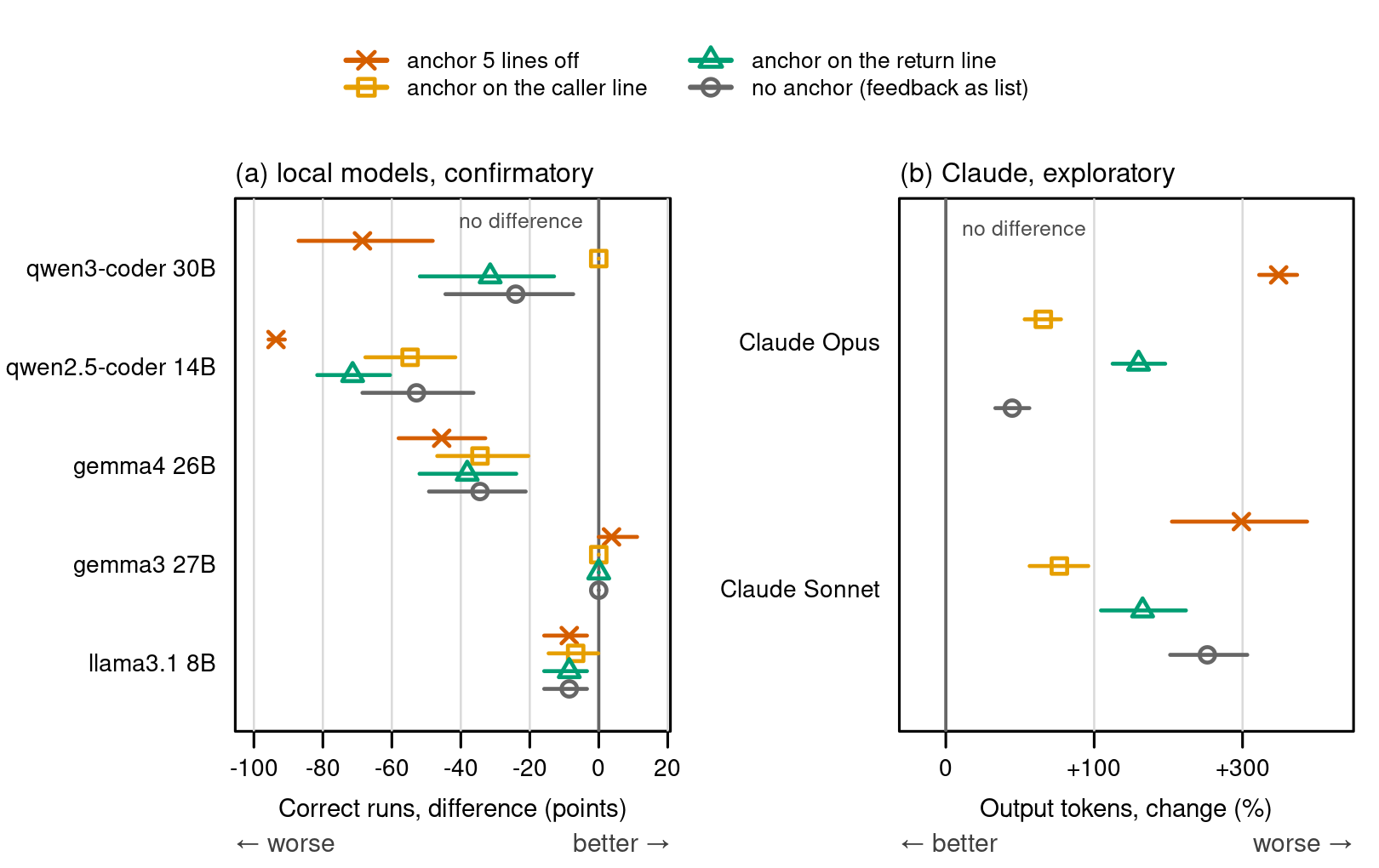}
  \caption{Misplaced-anchor experiment (T4--T6), typical-task effects relative to the accurate anchor (the zero line) with 95\% intervals: every wrong anchor lowered the local models' correctness, often to or below the level without any anchor (grey), and raised the Claude models' output tokens. Every misplaced arm keeps the treatment's wording.}
  \label{fig:misplaced}
\end{figure*}

Instead, the offset anchor cut qwen2.5-coder's correctness from 0.936 to 0.000, qwen3-coder's from 1.000 to 0.315, and gemma4's from 0.455 to 0.000, and all three have evidence to reject $H_0$ under both scorings (Holm $p \le 0.002$, Fig.~\ref{fig:misplaced}, Table~\ref{tab:newarms}). The return-line and caller-line anchors cost two or three of these models 31.5 to 71.4 points, with evidence to reject $H_0$ for each (Holm $p \le 0.047$), except that the caller line, which names the right function, left qwen3-coder unaffected. The nominal hierarchical checks do not support llama3.1's smaller losses (Appendix~\ref{app:hier}). 
The models trusted the quoted line and edited it, which produced wrong results, edit blocks that matched no line, or broken functions, and the offset anchor also tripled the typical answers of qwen2.5-coder and llama3.1. It was the only arm that damaged code that should not change: 16 of the 23 qwen2.5-coder runs that produced a script altered the result of a neighboring block. 
The Claude models (exploratory) show the same weakness in a different currency. Under the offset anchor, Opus stayed correct but generated 374\% more tokens. Sonnet's share of correct runs fell by 62.3 points under strict scoring and by 59.3 points under lenient scoring, and it damaged a neighboring block in 25 of 49 runs. 

\subsubsection{Contract Control}
\label{sec:contractctl}
In the main experiment the output contract changes with the tier, so the file-size effect of Section~\ref{sec:interaction} mixes size with contract. The contract-control experiment ran T1 to T3 under the edit-block contract, so that all six tiers share one contract. For each model we test the tier effect in Hypotheses~\eqref{eq:tier} under that contract, where $\theta_{\mathrm{T}j}$ is the mean task-level difference between treatment and control among the tasks of tier T$j$ under the edit-block contract. T1 to T3 come from the new runs and T4 to T6 from the main experiment.

Under that single contract, the correctness effect still depends on tier for the two coder models, which have evidence to reject $H_0$ under both scorings (permutation, Holm $p \le 0.0029$), and the token effect for llama3.1 (Holm $p = 0.00085$). The other local models and outcomes do not have evidence to reject $H_0$ (Holm $p \ge 0.11$). 

For the Claude models (exploratory, nominal $p$-values), the token savings still grow with tier under the single contract, and both have evidence to reject $H_0$ (permutation $p \le 0.00015$). 
The larger savings on large files are therefore not an artifact of the contract change. The contract matters at fixed size as well: on T4 under full rewrites, the treatment saved 15.8\% (Opus) and 21.8\% (Sonnet) of output tokens, against 38.9\% and 78.3\% under edit blocks (typical task, descriptive, six tasks). 
In this experiment the T1 to T3 differences come from new runs with 3 replicates per arm, and the T4 to T6 differences from the main experiment's runs with 13, made months earlier for Claude. The tiers therefore also differ in run period and precision (Section~\ref{sec:threats}).

\section{Discussion}

\textbf{Anchoring pays off as files grow.} On small files, finding the flawed line is easy and both formats saturate. As files grow, finding it becomes the challenge, and pointing to it has a large positive impact. The effect grows with file size even when every tier uses the same output contract (Section~\ref{sec:contractctl}). Thus, it is not an artifact of the contract change between T3 and T4. Tokens and cost cross over between the tier groups (Table~\ref{tab:cost}, Fig.~\ref{fig:main}c). On small files the export's longer request is a large share of a short exchange. Total tokens therefore rise for nearly every model, and the hosted cost of an edit rises for Opus. On large files the answer is long, and the export shortens it by more than it lengthens the request, so total tokens and cost both fall for the Claude models. The export's overhead grows with the number of comments, while the search it spares the model grows with the file, which is why the balance turns on the large files. For the local models the gain is in correctness rather than tokens, and it too arrives on the large tiers, where the model must find the flawed function among dozens before it can fix it.

\textbf{Show the model the right code.} A reviewer helps an AI editor most by showing it exactly which code each comment concerns, and harms it by showing the wrong code. With an accurate anchor and quote, the same requested changes made the strong models answer with far fewer tokens on large files and made the weaker models correct far more often (Sections~\ref{sec:tokens} and~\ref{sec:correctness}). The decomposition and misplaced-anchor experiments point to one mechanism (Sections~\ref{sec:decomp} and~\ref{sec:misplaced}): the models act on the pointer, above all on the quoted line, more than on the comment's wording. An accurate quote spares the model a search that a holistic list leaves to it, which a strong model turns into a shorter answer and a weaker model into a correct one. A wrong quote is followed, not corrected, even when the comment's text names the right code. Tools that consume anchored feedback should therefore check that each quoted line still occurs at its anchor, and reviewers should regenerate exports after the code changes.

\textbf{The human half of HAIC.} These results do not speak to fully autonomous ``vibe coding'' or ``agentic coding,'' where no human reviews the code at all. What the experiments quantify is the value of a reviewer's localized judgment once it is recorded \cite{bacchelli_expectations_2013, vaithilingam_expectation_2022, barke_grounded_2023}. The reviewer decides where the problem is and what should change, and the model does the editing. The format in which that judgment travels determines both what the editing costs and how often it succeeds, and the misplaced-anchor results show that the model takes the reviewer's pointer at its word. In this sense the experiment is evidence for keeping people in GAI-assisted development, because their targeted attention is worth both tokens and correctness, provided it points at the right code. The findings apply to review in which a person has already pinpointed each issue, and Section~\ref{sec:threats} discusses how they may transfer to real repositories and other languages.

\textbf{Model heterogeneity.} The models differ in whether they can act on anchored feedback. gemma3 never produced an edit block on the large files and llama3.1 mostly wrote malformed edit markers. Thus, their large-tier results measure the output format, not the feedback. gemma4 often spent its whole output budget on hidden reasoning, less often under the treatment. These differences are why we report each model separately rather than only pooled effects.

\textbf{Reviewer effort.} The output-token and cost savings accrue on the model's side, while anchoring asks the reviewer to select a line span for each comment. The keystroke-level model \cite{card_keystroke_1980} prices one anchored comment at about 5\,s. It takes a mental preparation (1.35\,s), two hand moves between keyboard and mouse ($2 \times 0.40$\,s), and pointing at both ends of the span ($2 \times 1.10$\,s). A click and a two-key shortcut add three keystrokes ($3 \times 0.28$\,s for an average non-secretary typist). That is 10 to 26\,s for the two to five comments per task in this benchmark, for a reviewer who has already found the flaw, which both formats require. A holistic comment may need extra words to say where the code is, so whether anchoring saves reviewer time overall is a question for a user study. The savings we measured are model-side efficiency, not workflow efficiency.

\textbf{Multi-turn exchanges.} The experiment prices one exchange, while practitioners often iterate \cite{barke_grounded_2023}. Across turns the effect could grow: in an agentic loop each turn's output becomes input to later turns, so a shorter edit also shrinks what they re-read, and a correct first edit removes a repair turn. It could also shrink, because an anchor refers to one version of a file, and an export that is not regenerated after an edit points at the wrong lines. The offset arm is a harsher version of that case, since it also moves the quoted lines, whereas a stale export still quotes the right code. Measuring these directions needs multi-turn runs, which we leave to future work.

\section{Threats to Validity}
\label{sec:threats}

\emph{Construct validity.} Correctness means passing validators that check the flawed blocks, and the collateral-damage check covers every other block of a library script. Total tokens count the task prompt with one tokenizer for every model and the output tokens as each provider reports them, so the hosted totals include hidden reasoning. The API view prices that tokenizer's prompt count, while Claude's own tokenizer may count more: the provider notes that its tokenizer for Claude 4.7 and later models produces about 30\% more tokens for the same text \cite{anthropic_pricing_2026}. With the prompt counted 1.5 times higher, the typical T4 to T6 edit via the API would still cost 16.4\% (Opus) and 39.3\% (Sonnet) less. The T1 to T3 premium would grow to 15.4\% and 2.6\%. Cost and latency are measured as run, and energy is not measured, so energy savings are an implication of fewer generated tokens rather than a result. 

\emph{Internal validity.} The main contrast changes wording and format together, and the decomposition experiment separates them. The output contract changes with tier, and the contract-control experiment separates contract from file size, although its small tiers come from new runs with 3 replicates per arm and its large tiers from the main experiment's runs with 13. Tier itself is a property of each task, not a randomly assigned factor, so the dependence of the effect on tier describes how the effect varies with file size rather than proving that size causes it. Hosted calls carried Claude Code's system prompt, its tool definitions, and the author's instruction files, identically in both arms. The new hosted arms ran three months after their references, across a CLI update and a change to those files. A drift check licenses that comparison but cannot exclude a confound as large as the effects compared, so those results are exploratory (Appendix~\ref{app:claude_r1}). In 26 of the 1{,}116 new hosted calls the model used tools, and excluding them changes no conclusion.

\emph{Statistical conclusion validity.} Task-level inference rests on 36 tasks, or 18 for the large tiers, so effects below the minimum detectable effects of Appendix~\ref{app:power} can go undetected, and per-tier results rest on six tasks and are descriptive (Section~\ref{sec:safeguards}). Holm correction runs within each family, and a correctness conclusion requires evidence under both scorings. The tasks of a tier are near-relatives (Appendix~\ref{app:padding}), so they may not be fully independent, which would make every task-level test somewhat optimistic. The pre-registered file-size test assumed equal variances across tiers, which does not hold, so a permutation test replaced it, and the two agree in every decision (Appendix~\ref{app:splitplot}). The hierarchical models, added after the results were known, violate their error assumptions even on the log scale and under the count model, so their $p$-values are nominal checks at best. Read that way, they reach the same decision as the task-level analysis in 80 of 89 comparisons and in every comparison of the main experiment. They do not support llama3.1's small new-arm correctness effects, which rest on a few tasks (Appendix~\ref{app:hier}). 

\emph{External validity.} The tasks are synthetic, single-file base-R programs, each exchange is a single turn, and the main experiment's anchors are exact. Every comment text names the code it concerns, so with vaguer comments a wrong anchor may hurt even more than the misplaced-anchor arms show. The seed is shown without line numbers, so the line-number arm measures numbers pasted into a chat, not an editor that displays them. Local models ran quantized on one 24\,GB GPU, and the original local runs were truncated to Ollama's default window, which the re-run corrects (Appendix~\ref{app:truncation}).

\emph{Transfer.} The misplaced-anchor results show that the reviewer's pinpointing must be right. The benchmark is synthetic base R with repeated structure. Library blocks come in five kinds, and 78 of the 110 single-line flaws have a look-alike line elsewhere in their file. Real repositories differ in ways that cut both ways. Their code is less repetitive, which makes content search easier for holistic feedback and could shrink the localization benefit. A real change, however, often spans several files, and an anchor records the file as well as the line, which a holistic comment must describe in words. Languages also differ in their edit units. Exact-match edit blocks may be more fragile under Python's significant indentation, Java methods map onto the function-patch contract, and markup or configuration files offer nothing to anchor to but lines. FileMark anchors any text file by line, so the delivery mechanism transfers, but the size of the effect must be measured in each setting, for example on repository benchmarks such as SWE-bench \cite{jimenez_swebench_2024}. Agent frameworks that plan and apply edits over several turns, rather than answering once through a command-line call, are a further setting we did not test.

\section{Conclusion}

Delivering a reviewer's requested changes anchored to the code they concern, with that code quoted, changed how GAI models edited it, and the benefit grew with the size of the file. In this paired experiment on synthetic base-R tasks, the export lengthened the request and shortened the answer. On small files the exchange used more tokens and, for Opus, cost more. On large files the Claude models' answers shrank enough to lower both the total tokens and the cost of an edit. Four of five local models made the requested changes correctly far more often. The decomposition placed the benefit in the format rather than the wording, and the misplaced-anchor experiment showed its critical assumption: the anchor must point at the right code. For HAIC in software engineering, the reviewer's pointer is the part of the feedback that has an outsized impact. Future work includes real repositories, for example a subset of SWE-bench, other languages, agent frameworks that apply the edits themselves, multi-turn exchanges with stale exports, and a user study of the reviewer's effort.

\section*{Data and Code Availability}

All templates, generated task instances, harness and analysis code, raw model outputs, results, and this paper's source are available at \url{https://github.com/billyl320/filemark_usage}. The original study comprises 9{,}828 runs executed from June 25 to July 7, 2026: the 6{,}552-run main experiment and the 3{,}276-run exploratory function-patch wave, including 142 re-executed Sonnet runs whose superseded responses are kept. From October 9 to 14, 2026, the revision re-ran 4{,}836 local runs with the corrected context window and kept the truncated originals. It also added 2{,}610 local and 1{,}044 hosted runs of the new arms and a 72-call hosted drift check. A SHA-256 manifest shows that no pre-revision file changed, and an append-only log records every revision step. The repository also contains the FileMark \texttt{.feedback.md} review exports produced while revising this paper (\texttt{old\_feedback/}). The manuscript was itself edited through the tool under study, and those artifacts document each revision round.

\section*{Acknowledgments}
The following tools were used in this work: R, Ollama, Claude Code via VSCodium, FileMark, \LaTeX{}, VSCodium, git, GitHub, Linux terminal commands (e.g., grep, cd, etc.), and Ubuntu Linux. Claude (Opus 4.8 and Fable 5 for the original study, and Opus 5.5 and Fable 5.1 for this revision) assisted with experiment orchestration, analysis code, and manuscript preparation. Tools not mentioned were unintentionally omitted. The manuscript was itself revised using the feedback format under study: the author recorded review comments on each draft as FileMark feedback files and provided them to the AI assistant when requesting revisions. Those exports are preserved in the repository. The author is the developer of the FileMark extension evaluated in this study. The reviewers provided substantive feedback during the review process, and their suggestions have strengthened the paper. We thank the Department of Computational and Data Sciences at George Mason University for partially funding this effort.

\bibliographystyle{IEEEtran}
\bibliography{main_r1}

\appendices

\section{Statistical Methods}
\label{app:stats}

This appendix explains each method of the analysis, what it assumes, and why it fits this design. All computations use R 4.5.2 \cite{rcore_r_2025}, and the scripts are in the repository.

\subsection{Two-Stage Task-Level Analysis}
\label{app:twostage}
Each model ran each task 13 times per arm in the main experiment and 3 times in a new arm. These replicates share the task's content, so their outcomes are correlated (Appendix~\ref{app:power}), and counting them as independent observations would overstate the evidence. We therefore use a summary-measures analysis \cite{matthews_serial_1990}, which reduces the repeated observations of each unit to one summary and then analyzes the summaries. In the first stage, the runs of model $m$, task $i$, and arm $a$ are reduced to the share of correct runs $\bar{c}_{mia}$ and, for each token, cost, and time outcome, the mean log value $\bar{\ell}_{mia}$. Here $\ell = \log(\text{count} + 1)$ for tokens and $\ell = \log(\text{value})$ for cost and time. In the second stage, each task contributes one paired difference between the arms, $d_i = \bar{c}_{miT} - \bar{c}_{miC}$ for correctness and $d_i = \bar{\ell}_{miT} - \bar{\ell}_{miC}$ for the log outcomes. All inference uses these $n$ differences ($n = 36$ in the main experiment and 18 in the new arms). The tasks are thus treated as a sample from a population of tasks, as the items of a language study are \cite{clark_language_1973}.

For the log outcomes we report two estimands. The \emph{typical-task} change is $\exp(\bar{d}) - 1$, the back-transformed mean of the task-level differences: the geometric mean over tasks of the ratio of the arms' geometric-mean values. It is the quantity the sign-flip test addresses. The \emph{aggregate} change is $\hat{\gamma} - 1$, with $\hat{\gamma} = \sum_i \bar{y}_{miT} / \sum_i \bar{y}_{miC}$ the ratio of the arms' totals of the task means $\bar{y}$ on the original scale, which weights long tasks in proportion to their size and is the accounting of a whole workload. Its test, added in the revision, swaps the two arms within tasks (Hypotheses~\eqref{eq:agg}). Under the null hypothesis of no effect in any task, the pair $(\bar{y}_{miT}, \bar{y}_{miC})$ is exchangeable, so the $p$-value is the share of the $2^n$ arm assignments whose $|\log|$ of the ratio of totals is at least the observed one. The assignments are enumerated for $n \le 20$ and drawn 100{,}000 times for $n = 36$. The two estimands differ in sign for total tokens over all 36 tasks for Opus and qwen3-coder, because the treatment adds tokens on small files and removes them on large ones.

\subsection{Sign-Flip Permutation Test}
For each model and contrast, the test concerns the arm contrast in Hypotheses~\eqref{eq:arm},
where $\theta$ is the mean of the task-level differences $d_i$ in the population of tasks. Under $H_0$ the differences are taken to be symmetric around zero, so each $d_i$ is as likely to be positive as negative, and flipping the sign of any $d_i$ gives an equally likely data set. The test statistic is the absolute mean difference $|\bar{d}|$, and the $p$-value is the share of the sign patterns whose statistic is at least the observed one \cite{good_permutation_2005}:
\[ p = \frac{1}{2^n} \sum_{s} \mathbf{1}\Bigl\{ \Bigl| \frac{1}{n} \sum_{i=1}^{n} s_i d_i \Bigr| \ge |\bar{d}| \Bigr\}, \]
where:
\begin{itemize}
  \item $n$: the number of tasks.
  \item $d_i$: the task-level difference of task $i$, and $\bar{d}$ their mean.
  \item $s = (s_1, \ldots, s_n)$: a sign pattern with each $s_i$ either $-1$ or $+1$, summed over all $2^n$ patterns.
  \item $\mathbf{1}\{\cdot\}$: 1 when the condition holds and 0 otherwise.
\end{itemize}
We enumerate all patterns for $n \le 20$, which makes the test exact, and draw 100{,}000 random patterns for $n = 36$, counting the observed pattern among them. The test assumes only the symmetry that $H_0$ implies for paired differences, so it remains valid for the bounded and often tied correctness differences. The Wilcoxon signed-rank test \cite{wilcoxon_individual_1945} of the same differences repeats each test with ranks. Repeating a test under different assumptions is a sensitivity analysis, and a conclusion that holds under both is more credible \cite{thabane_sensitivity_2013}.

\subsection{Cluster Bootstrap Intervals}
Intervals resample whole tasks rather than runs, so they keep the dependence among a task's replicates \cite{field_bootstrapping_2007}. Each of 10{,}000 bootstrap samples draws, within each tier, as many tasks as the tier has, with replacement, and recomputes the effect. The 95\% interval runs from the 2.5th to the 97.5th percentile of the bootstrap effects. Resampling within tiers keeps every sample's mix of file sizes equal to the design's. The intervals are not adjusted for multiple tests, so an interval can exclude zero while the Holm-adjusted test gives no evidence to reject the null hypothesis.

\subsection{Split-Plot Analysis of Variance}
\label{app:splitplot}
A split-plot design has two kinds of experimental units: whole plots receive the levels of one factor, and the split plots within each whole plot receive the levels of a second. In Montgomery's paper-mill example \cite{montgomery_design_2020}, each batch of pulp is made by one of three preparation methods, so the batch is the whole plot, and it is divided into samples cooked at four temperatures, the split plots. The method is therefore tested against the variation between batches and the temperature against the variation within them.

In our design a task is the whole plot, its tier is the whole-plot factor, and the two arms are the split plots within the task. The arm $\times$ tier interaction asks whether the arm effect differs between tiers, the tier effect in Hypotheses~\eqref{eq:tier},
where $\theta_{\mathrm{T}j}$ is the mean task-level difference between treatment and control in tier T$j$. With two arms, the interaction $F$ test of the analysis plan (5 and 30 degrees of freedom) is the same test as a one-way ANOVA of the 36 task-level differences across the six tiers. It therefore uses one value per task, not the 6{,}552 runs of our first analysis. In the paper mill the whole-plot factor, the preparation method, was assigned to batches at random. Our whole-plot factor, the tier, was not assigned: it is a property of each task, so the interaction describes how the effect varies with file size rather than proving that size causes it.

The $F$ test assumes that the differences are normal with the same variance in every tier. The Fligner--Killeen test of the second assumption concerns
\begin{equation}\label{eq:var}
\begin{aligned}
  H_0 &: \sigma^2_{\mathrm{T1}} = \sigma^2_{\mathrm{T2}} = \cdots = \sigma^2_{\mathrm{T6}} \\
  H_1 &: \sigma^2_{\mathrm{T}j} \neq \sigma^2_{\mathrm{T}k} \text{ for at least one pair of tiers } j, k
\end{aligned}
\end{equation}
where $\sigma^2_{\mathrm{T}j}$ is the variance of the task-level differences in tier T$j$. Across tiers, the standard deviation of the pooled differences ranges from 0.025 to 0.152 for log tokens and from 0.013 to 0.126 for correctness, and for this test we have evidence to reject $H_0$ (Fligner--Killeen $p \le 0.0063$). Because that assumption fails, the primary test is a permutation test of the tier effect, a deviation from the analysis plan made during the revision \cite{good_permutation_2005}. The tier labels of the task-level differences are reassigned at random 100{,}000 times, and the one-way $F$ statistic is recomputed each time. The $p$-value is the share of reassignments with a statistic at least the observed one, the observed one counted among them. Under $H_0$ the differences are exchangeable across tiers, so the test needs no distributional assumption, and its smallest attainable $p$ is $10^{-5}$. The $F$ test, Welch's ANOVA, which allows unequal variances \cite{welch_comparison_1951}, and the Kruskal--Wallis test of the differences \cite{kruskal_ranks_1952} are reported as nominal checks. For the pooled data, all four give evidence to reject $H_0$ for output tokens and for correctness under both scorings ($p \le 2 \times 10^{-4}$). 
For the 34 per-model tests with a Holm decision, the $F$ and Kruskal--Wallis tests reach the same decision as the permutation test in every case, and Welch's ANOVA does so in all 15 cases in which it is defined. It is undefined when a tier has no variation at all, as for a model correct in every run of one tier. 

\subsection{Hierarchical Models of the Individual Runs}
\label{app:mixed}
A hierarchical model of the individual runs uses every run and estimates the variation between tasks \cite{baayen_mixed_2008}. For one model and a contrast between arms $A$ and $B$, every model below has the same structure. Run $r$ of task $i$ has an arm indicator $x_{ir}$ (1 for arm $A$, 0 for arm $B$). Its expected outcome, written on the scale of the model, is the sum of a fixed intercept and arm effect and of task $i$'s own deviations from both. This linear predictor is
\[ \eta_{ir} = (\beta_0 + u_{0i}) + (\beta_1 + u_{1i})\,x_{ir}, \]
where:
\begin{itemize}
  \item $\eta_{ir}$: the expected outcome of run $r$ of task $i$ on the model's scale. That scale is the log of the mean token count for the count model, the expected log value for the linear model, and the probability of a correct run for the linear probability model.
  \item $x_{ir}$: the arm indicator, 1 for arm $A$ and 0 for arm $B$.
  \item $\beta_0$: the mean of the linear predictor under arm $B$.
  \item $\beta_1$: the average arm effect, the parameter of interest.
  \item $u_{0i}$ and $u_{1i}$: task $i$'s deviations from $\beta_0$ and $\beta_1$, normal with mean zero and covariance matrix $\Sigma$.
\end{itemize}
The random arm effect $u_{1i}$ stays in every fit. Without it, the arm effect would be tested against the variation between runs alone, which is the pseudo-replication the task-level analysis avoids \cite{barr_random_2013}. Every fit uses \texttt{lme4} \cite{bates_lme4_2015}, and tests
\begin{equation}\label{eq:hier}
\begin{aligned}
  H_0 &: \beta_1 = 0 \\
  H_1 &: \beta_1 \neq 0
\end{aligned}
\end{equation}
where $\beta_1$ is the average arm effect defined above.

\emph{Token counts.} Output and total tokens are counts, so the reported model is a negative-binomial count model with mean $\exp(\eta_{ir})$, the mean count implied by the linear predictor defined above, and dispersion parameter $\theta$. Its arm effect $\exp(\beta_1) - 1$ is the multiplicative change in the mean count, the same kind of quantity as the typical-task change. A Poisson model is far too dispersed for these counts. Its Pearson dispersion ratio, which is 1 when the model holds, lies between 1.4 and 519 for the main experiment's output tokens and reaches 1{,}304 in the new arms. The negative binomial's lies between 0.6 and 2.1 \cite{hilbe_negative_2011}. The fit uses the Laplace approximation and a Wald $z$ test. For the 9 new-arm contrasts in which it did not converge and the 1 in which it was singular, we report instead a quasi-Poisson model with a separate intercept for each task and CR2 cluster-robust standard errors \cite{pustejovsky_small_2018, verhoef_quasipoisson_2007}. Its Holm-adjusted $p$-value is then the one compared with the task-level decision in Appendix~\ref{app:hier}, and the same model is also reported for every contrast as a check. The linear mixed model on $\log(\text{count} + 1)$ with a Kenward--Roger $F$ test \cite{kenward_small_1997, halekoh_kenward_2014} is a further check. 

\emph{Correctness.} The outcome is 1 for a correct run and 0 otherwise. A logistic mixed model is the usual choice, but it cannot be estimated here. In 81.9\% of the main experiment's task-and-arm cells, every run of a model was correct or none was, so the logistic likelihood has no finite maximum for those tasks (separation). The reported model is therefore a linear probability model \cite{gomila_logistic_2021}: the linear mixed model with $\eta_{ir}$ as the probability of a correct run and a normal error term, fitted by restricted maximum likelihood and tested with the Kenward--Roger $F$ test. Its $\beta_1$ is the difference in the probability of a correct run, in the same units as the task-level difference. Its residuals take two values at each fitted value, so a cluster-robust check with no distributional assumption repeats every comparison: ordinary least squares with a separate intercept for each task, $y_{ir} = \alpha_i + \beta_1 x_{ir} + e_{ir}$, with CR2 standard errors by task \cite{pustejovsky_small_2018}. 

Appendix~\ref{app:hier} shows that the error assumptions of these models fail, so their $p$-values are nominal checks.

\subsection{Intraclass Correlation and Design Effect}
The intraclass correlation $\rho$ is the share of the variance of the paired differences that lies between tasks rather than between the replicates of one task \cite{killip_icc_2004}. With $m$ replicates per task, the design effect $1 + (m - 1)\rho$ is the factor by which this clustering inflates the variance of a mean, compared with independent observations. A design of $n$ tasks with $m$ replicate pairs each therefore carries about as much information as $nm / (1 + (m - 1)\rho)$ independent pairs (Appendix~\ref{app:power}).

\subsection{Minimum Detectable Effects}
Post hoc power adds nothing to the observed $p$-values \cite{hoenig_abuse_2001}, so we report minimum detectable effects (MDEs) instead: the smallest true effect that the design detects with power 0.80 in a two-sided test at $\alpha = 0.05$ \cite{bloom_mde_1995}. Their standard deviation is the one observed in the experiment, so they are a sensitivity analysis of the design made after the fact, not a planned power (Appendix~\ref{app:power}). For $n$ task-level differences with standard deviation $s$,
\[ \mathrm{MDE} = \left(t_{0.975,\,n-1} + t_{0.80,\,n-1}\right) \frac{s}{\sqrt{n}}, \]
where:
\begin{itemize}
  \item $n$ and $s$: the number of task-level differences and their standard deviation.
  \item $t_{q,\,n-1}$: the $q$ quantile of the $t$ distribution with $n - 1$ degrees of freedom, $q = 0.975$ for the two-sided test and $q = 0.80$ for the power.
\end{itemize}

\subsection{The Holm--Bonferroni Procedure}
\label{app:holm}
For each family of related tests we control the familywise error rate at $\alpha = 0.05$ with the Holm--Bonferroni step-down procedure \cite{holm_simple_1979}, computed by base R's \texttt{p.adjust}. Let $m$ denote the number of tests in the family and $p_{(1)} \le p_{(2)} \le \cdots \le p_{(m)}$ the family's $p$-values in ascending order. The procedure rejects hypotheses sequentially, starting from $p_{(1)}$, for as long as
\[ p_{(k)} \le \frac{\alpha}{m - k + 1}, \]
and stops at the first $k$ for which the inequality fails. That hypothesis and all remaining ones are retained. The families are as follows. In the main experiment and in the function-patch wave, each outcome and scoring forms one family across the seven models, and a model with no task-level difference (such as Claude Opus at ceiling) has no test. In the new arms, each experiment and outcome forms one family across the five local models, with all of the experiment's named contrasts together: 10 tests for the decomposition, 15 for the misplaced anchors, and 5 for the contract control. The hierarchical models use the same families. Adjusted values are reported as $p_H$, and we have evidence to reject the null hypothesis of a test only where $p_H < 0.05$.

\section{Hierarchical Robustness Check}
\label{app:hier}

The hierarchical models of Appendix~\ref{app:mixed} were fitted for every model in the main experiment (output tokens, total tokens, and correctness) and for every confirmatory contrast of the decomposition and misplaced-anchor experiments (output tokens and correctness). That gives 39 count fits and 50 correctness fits, after 14 correctness contrasts in which every run or no run was correct left nothing to model. Each fit tests Hypotheses~\eqref{eq:hier},
where $\beta_1$ is the average arm effect of Appendix~\ref{app:mixed}. Each fit is tested once with the reported model and once with the cluster-robust check. The reported model is the negative-binomial count model for tokens, or the backup where it did not converge, and the linear probability model for correctness. The check is the quasi-Poisson model for tokens and ordinary least squares for correctness. 

Table~\ref{tab:hier} compares the analyses for the main experiment, and Table~\ref{tab:hier_agree} counts the agreement in every experiment. Read as nominal checks, the reported models reach the same Holm decision as the task-level analysis in 80 of the 89 comparisons that at least one of them can test, including all 26 of the main experiment. They point the same way in 88 of 89. The cluster-robust check agrees in 73 of 89. Its standard errors for the long-tailed counts are wide, so it does not confirm the Claude token effects that the task-level test, the count model, and the aggregate permutation test all find. The log linear mixed model agrees in all 38 token comparisons it could fit, and its fit failed for one. 

\begin{table*}[!t]
\caption{Main experiment: the task-level typical-task analysis against the reported hierarchical model (negative binomial for tokens, linear probability model for correctness) and the cluster-robust check (quasi-Poisson for tokens, least squares for correctness), with nominal Holm $p$-values for the two checks. Bold with an asterisk marks evidence to reject the null hypothesis of no difference, underlined where the treatment did better.}
\label{tab:hier}
\centering
\begin{fittable}
\begin{tabular}{llrrrrr}
\toprule
 & & \multicolumn{2}{c}{Task level} & \multicolumn{2}{c}{Hierarchical model} & Robust check \\
\cmidrule(lr){3-4}\cmidrule(lr){5-6}\cmidrule(lr){7-7}
Model & Outcome & Effect & $p_H$ & Effect & $p_H$ & $p_H$ \\
\midrule
Claude Opus & Total tokens (\%) & \sig{$+6.9$} & \sigp{0.047} & \sig{$+6.8$} & \sigp{0.032} & 0.67 \\
 & Output tokens (\%) & \good{$-14.7$} & \sigp{0.014} & \good{$-13.9$} & \sigp{0.011} & 0.25 \\
 & Correct, lenient (points) & 0.0 & n/a & n/a & n/a & n/a \\
\addlinespace
Claude Sonnet & Total tokens (\%) & $-4.2$ & 0.62 & $-4.6$ & 0.49 & 0.23 \\
 & Output tokens (\%) & \good{$-50.5$} & \sigp{$<10^{-4}$} & \good{$-52.5$} & \sigp{$<10^{-4}$} & 0.25 \\
 & Correct, lenient (points) & $-0.9$ & 1 & $-0.9$ & 0.65 & 0.65 \\
\addlinespace
qwen3-coder 30B & Total tokens (\%) & $+4.4$ & 0.78 & $+3.7$ & 0.89 & 0.67 \\
 & Output tokens (\%) & $-16.7$ & 0.13 & $-22.3$ & 0.052 & 0.18 \\
 & Correct, lenient (points) & \good{$+25.9$} & \sigp{0.0062} & \good{$+25.9$} & \sigp{0.0032} & \sigp{0.0031} \\
\addlinespace
qwen2.5-coder 14B & Total tokens (\%) & \sig{$+12.2$} & \sigp{$<10^{-4}$} & \sig{$+12.0$} & \sigp{$<10^{-4}$} & 0.21 \\
 & Output tokens (\%) & $-7.9$ & 0.28 & $-10.4$ & 0.063 & 0.25 \\
 & Correct, lenient (points) & \good{$+30.6$} & \sigp{0.0002} & \good{$+30.6$} & \sigp{0.0004} & \sigp{0.0004} \\
\addlinespace
gemma4 26B & Total tokens (\%) & $-2.1$ & 0.78 & $-2.1$ & 0.89 & 0.077 \\
 & Output tokens (\%) & $-10.1$ & 0.053 & $-9.0$ & 0.052 & \sigp{0.029} \\
 & Correct, lenient (points) & \good{$+20.6$} & \sigp{0.0004} & \good{$+20.5$} & \sigp{0.0009} & \sigp{0.0009} \\
\addlinespace
llama3.1 8B & Total tokens (\%) & \sig{$+14.7$} & \sigp{$<10^{-4}$} & \sig{$+15.6$} & \sigp{$<10^{-4}$} & 0.67 \\
 & Output tokens (\%) & $-7.7$ & 0.73 & $-4.4$ & 1 & 0.94 \\
 & Correct, lenient (points) & \good{$+9.2$} & \sigp{0.0002} & \good{$+9.2$} & \sigp{0.0016} & \sigp{0.0016} \\
\addlinespace
gemma3 27B & Total tokens (\%) & \sig{$+11.4$} & \sigp{$<10^{-4}$} & \sig{$+11.2$} & \sigp{$<10^{-4}$} & \sigp{0.0061} \\
 & Output tokens (\%) & $-0.4$ & 0.81 & $-0.6$ & 1 & 0.57 \\
 & Correct, lenient (points) & $-0.9$ & 1 & $-0.9$ & 0.65 & 0.65 \\
\bottomrule
\end{tabular}
\end{fittable}
\end{table*}

\begin{table*}[!t]
\caption{Decomposition, misplaced-anchor, and main experiments: agreement of the hierarchical models with the task-level analysis, counted over the comparisons that at least one analysis can test; the log linear mixed model is counted on the token comparisons only; the last columns count the token contrasts fitted by the negative-binomial model and those that used the backup.}
\label{tab:hier_agree}
\centering
\begin{fittable}
\begin{tabular}{lrrrrrrrrr}
\toprule
 & \multicolumn{3}{c}{Reported model} & \multicolumn{3}{c}{Robust check} & Log LMM & \multicolumn{2}{c}{Count fits} \\
\cmidrule(lr){2-4}\cmidrule(lr){5-7}\cmidrule(lr){8-8}\cmidrule(lr){9-10}
Experiment & Tests & \hd{Same}{decision} & \hd{Same}{direction} & Tests & \hd{Same}{decision} & \hd{Same}{direction} & \hd{Same}{decision} & \hd{Negative}{binomial} & Backup \\
\midrule
Main experiment & 26 & 26 & 26 & 26 & 20 & 22 & 14 of 14 & 14 of 14 & 0 \\
Decomposition & 24 & 21 & 24 & 24 & 21 & 23 & 10 of 10 & 8 of 10 & 2 \\
Misplaced anchors & 39 & 33 & 38 & 39 & 32 & 38 & 14 of 14 & 7 of 15 & 8 \\
\midrule
All & 89 & 80 & 88 & 89 & 73 & 83 & 38 of 38 & 29 of 39 & 10 \\
\bottomrule
\end{tabular}
\end{fittable}
\end{table*}

Nine comparisons reach a different decision. Six are llama3.1's correctness effects of 8.5 points in three new-arm contrasts, each under both scorings. llama3.1 was correct in no run of one arm and in 8.5\% of the runs of the other, all on 9 of the 18 tasks. The sign-flip test sees nine task differences in one direction, whereas the models also weigh how much the effect varies between tasks. We therefore do not have strong evidence for llama3.1's new-arm correctness effects. The other three are new-arm token contrasts in which the count model did not converge and the less sensitive backup does not confirm the task-level evidence: gemma4's format and return-line effects and llama3.1's offset-anchor effect. 

\subsection{Model Assumptions and the Scale of the Token Counts}
\begin{figure*}[!t]
  \centering
  \includegraphics[width=0.8\textwidth]{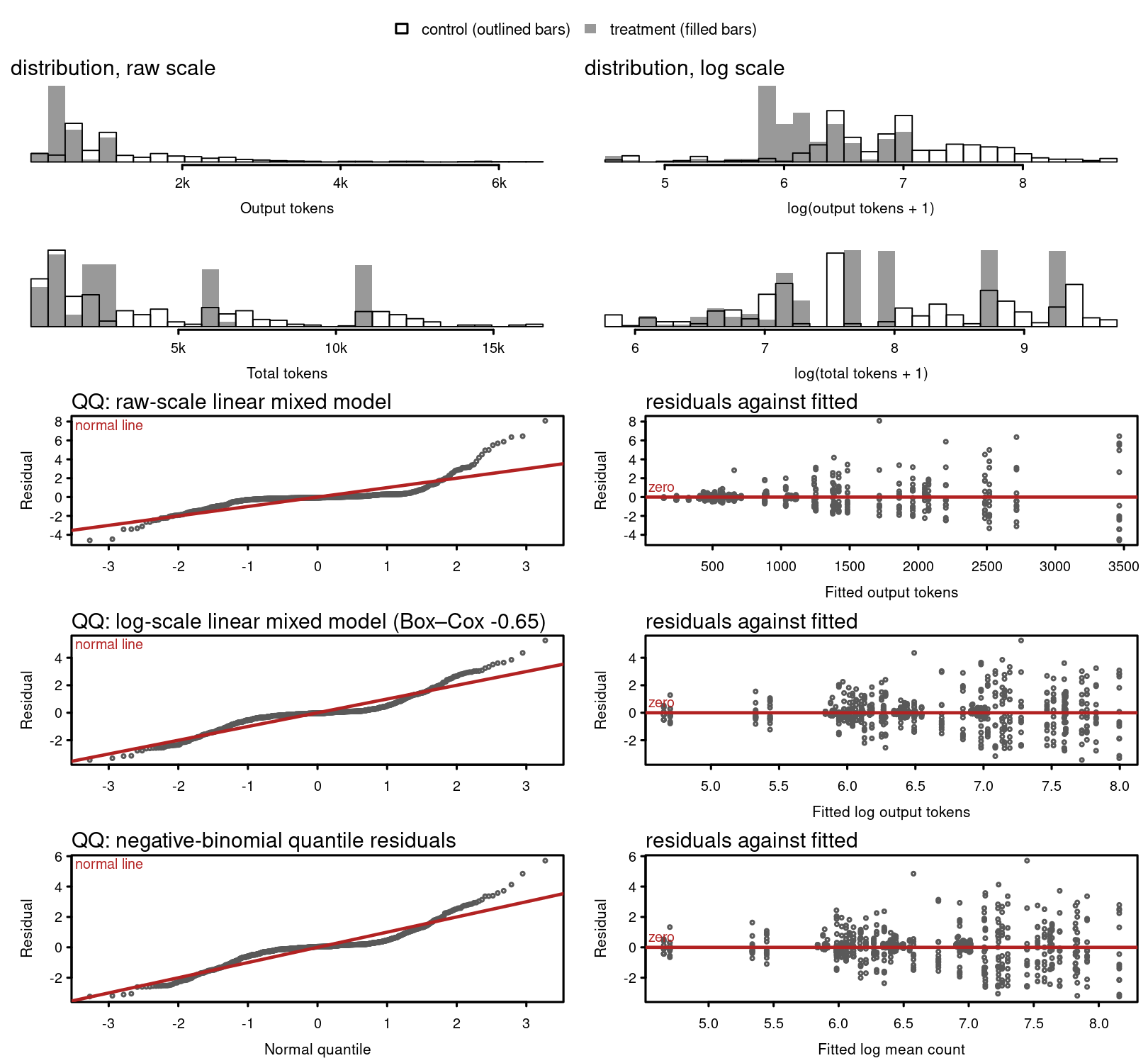}
  \caption{Main experiment, Claude Sonnet as a representative model: output and total tokens on the raw and the log scale (top rows), and residual diagnostics of three hierarchical models. The residuals keep heavy tails and a spread that grows with the fitted value, and the repository draws these panels for every model.}
  \label{fig:hier_sonnet}
\end{figure*}

The models make four assumptions, which we checked as follows.
\begin{enumerate}
  \item Linearity: every model has one binary predictor, the arm, so the arm effect is a difference of two means (on the link scale) and the linear form holds by construction.
  \item Independent predictors: with a single predictor there is nothing for it to be correlated with.
  \item The error distribution: this assumption fails. Figure~\ref{fig:hier_sonnet} shows, for Claude Sonnet as a representative model, the outcome itself on the raw and the log scale and the residual diagnostics of three models: a linear mixed model on the raw scale, the same model on the log scale, and the negative-binomial model's randomized quantile residuals, which are standard normal when the model holds \cite{dunn_randomized_1996}. The script \texttt{R/figures\_r1.R} in the public repository (Data and Code Availability) draws the same panels for the other six models. Across the seven models, the raw residuals have a skewness of magnitude up to 7.7 and an excess kurtosis up to 86, and the log lowers the skewness for every model but gemma4. The Box--Cox family \cite{box_analysis_1964}, which includes the log as the exponent 0, estimates exponents between $-0.65$ and 0.55, so the log is a reasonable single transformation and the one that keeps effects interpretable as percentage changes. Despite the log and the count model, the residuals keep heavy tails, with an excess kurtosis up to 53 on the log scale, and their spread still changes with the fitted value for five of the seven models (Spearman correlations of the absolute residual with the fitted value of magnitude 0.17 to 0.43). For Sonnet it still grows, so the errors look multiplicative even after the transformation, and the $p$-values of these models are nominal checks at best. 
  \item Independent errors: runs of one task share the task deviations, which every model represents, and runs of different tasks are assumed independent, which the near-relative tasks of Appendix~\ref{app:padding} may strain.
\end{enumerate}
For correctness the residuals take only two values at each fitted value, so the third assumption fails by construction. None of the paper's conclusions rests on these models, because the task-level permutation tests make no distributional assumption.

\subsection{Fits That Did Not Converge or Lie on a Boundary}
Of the 39 count fits, 30 converged, one of them to a singular fit, and 9 failed, all in the new arms, so the backup gave the decision in those 10. The failed fits are the sparsest contrasts, 3 replicates against 13 on 18 tasks, for gemma4, llama3.1, and qwen3-coder, whose answers in some arms were almost always cut at the output limit or empty. Of the 89 linear mixed-model fits, 42 are singular, mostly because many tasks have identical outcomes in both arms. In 28 of them the correlation between a task's intercept and its arm deviation is estimated at $\pm 1$, and in 14 a variance is estimated at zero. The arm effect is still estimated in a singular fit, and its decision matched the cluster-robust check in every one. An earlier version of this check used a weak penalty on the covariance, which is a prior in Bayesian terms, and is kept in the repository but not reported. 

\section{Main Experiment by Tier}
\label{app:tiers}

Table~\ref{tab:tiers_full} breaks the main experiment's results of Tables~\ref{tab:main} and~\ref{tab:correct} down by model and tier, as the aggregate change in each tier's six tasks (descriptive, with the limits of per-tier inference in Section~\ref{sec:safeguards}). Total tokens rise on T1 for every model, by 19\% to 38\%, because the longer request is a large share of a short exchange. They fall on T4 for the Claude models, qwen3-coder, and gemma4, where the shorter answer outweighs it. Token changes for a model that fails on the format track the length of unusable output. llama3.1's $+250.8\%$ in output tokens at T4, for example, is longer malformed output, not more editing. 

\begin{table*}[!t]
\caption{Main experiment by size tier: Tables~\ref{tab:main} and~\ref{tab:correct} broken down by model and tier (descriptive, six tasks per cell), as treatment minus control in total tokens and output tokens (aggregate \% change) and in correct runs (lenient scoring, points).}
\label{tab:tiers_full}
\centering
\begin{fittable}
\begin{tabular}{lrrr>{\columncolor{tablegray}}r>{\columncolor{tablegray}}r>{\columncolor{tablegray}}r}
\toprule
Model & T1 & T2 & T3 & T4 & T5 & T6 \\
\midrule
\multicolumn{7}{l}{\emph{Total tokens, aggregate change (\%)}} \\
Claude Opus & $+32.3$ & $+15.8$ & $+13.0$ & $-10.9$ & $-1.8$ & $-4.6$ \\
Claude Sonnet & $+18.8$ & $+9.5$ & $+6.7$ & $-32.8$ & $-12.5$ & $-10.8$ \\
\addlinespace
qwen3-coder 30B & $+33.8$ & $+13.6$ & $+12.1$ & $-37.5$ & $+0.8$ & $-0.1$ \\
qwen2.5-coder 14B & $+36.0$ & $+13.6$ & $+12.3$ & $+8.6$ & $+3.0$ & $+1.9$ \\
gemma4 26B & $+28.2$ & $-3.6$ & $-12.0$ & $-11.6$ & $-7.5$ & $-3.8$ \\
llama3.1 8B & $+37.8$ & $+14.2$ & $+13.5$ & $+34.6$ & $+2.0$ & $-1.2$ \\
gemma3 27B & $+32.9$ & $+12.7$ & $+11.7$ & $+5.6$ & $+5.0$ & $+1.6$ \\
\midrule
\multicolumn{7}{l}{\emph{Output tokens, aggregate change (\%)}} \\
Claude Opus & $+11.8$ & $+8.9$ & $+6.2$ & $-38.4$ & $-24.3$ & $-37.0$ \\
Claude Sonnet & $-8.8$ & $-3.0$ & $-5.2$ & $-80.3$ & $-74.7$ & $-76.5$ \\
\addlinespace
qwen3-coder 30B & $-8.9$ & 0 & $-0.5$ & $-75.2$ & $-26.3$ & $-26.9$ \\
qwen2.5-coder 14B & $+0.8$ & 0 & $-0.1$ & $-16.9$ & $-27.1$ & $-29.0$ \\
gemma4 26B & $+23.8$ & $-9.7$ & $-19.1$ & $-17.7$ & $-16.0$ & $-12.2$ \\
llama3.1 8B & $+0.6$ & $+1.0$ & $+2.2$ & $+250.8$ & $-22.6$ & $-49.0$ \\
gemma3 27B & $-10.6$ & $-0.5$ & $+0.2$ & $-0.1$ & $+5.5$ & 0 \\
\midrule
\multicolumn{7}{l}{\emph{Correct runs, difference (lenient scoring, points)}} \\
Claude Opus & 0 & 0 & 0 & 0 & 0 & 0 \\
Claude Sonnet & $-5.1$ & 0 & 0 & 0 & 0 & 0 \\
\addlinespace
qwen3-coder 30B & 0 & 0 & 0 & $+23.1$ & $+83.3$ & $+48.7$ \\
qwen2.5-coder 14B & $+1.3$ & 0 & 0 & $+34.6$ & $+73.1$ & $+74.4$ \\
gemma4 26B & 0 & 0 & $+3.8$ & $+37.2$ & $+50.0$ & $+32.8$ \\
llama3.1 8B & $+10.3$ & $+3.8$ & $+15.4$ & $+1.3$ & $+12.8$ & $+11.5$ \\
gemma3 27B & $-5.1$ & 0 & 0 & 0 & 0 & 0 \\
\bottomrule
\end{tabular}
\end{fittable}
\end{table*}

\section{Local Context Truncation in the Original Runs}
\label{app:truncation}
The original local runs did not set Ollama's context window, so it stayed at the default of 4{,}096 tokens. Every one of the 3{,}117 local T5 and T6 runs reports exactly 4{,}096 prompt tokens. The full prompts, recovered from the token arrays Ollama returns, run from about 5{,}300 to 12{,}000 tokens. Ollama keeps the first few tokens and the end of an over-long prompt, so the instructions, the output contract, and most of the script were lost, while the requested changes at the end survived. The loss was not neutral between the arms, because the treatment's quoted lines sit in the surviving end. On T1 to T4 every prompt fit, but the window must also hold the answer, so an answer that outgrew the remaining space was cut. This happened in 637 runs. gemma4 was cut in every tier: 8 of 156 runs on T1, 23 on T2, 96 on T3, and 307 of 310 on T4. On T4, gemma3 was cut in 156 of 312 runs, qwen3-coder in 39, and llama3.1 in 8. qwen2.5-coder was never cut. In all, 3{,}754 of the 7{,}020 local runs were affected. 

A pilot with a 32{,}768-token window confirmed that every local model then reads the full prompt. On the longest prompt, \texttt{qwen3-coder:30b} counts 10{,}304 tokens, exactly the length recovered from the original run's token array, and each model ran entirely on the GPU. Re-running five archived small-tier runs per arm under the original settings reproduced all 10 responses byte-for-byte, so the weights, runtime, and seeded sampling are unchanged, and unaffected cells could be kept.

Table~\ref{tab:truncation} compares the local results of the main experiment before and after the correction. The truncation hid most of the treatment's correctness effect. In the original data only llama3.1's gain of 5.1 points survived Holm correction. In the corrected data four of the five local models gain, by 9.2 to 30.6 points. The token picture changed as well, in aggregate: qwen2.5-coder's $+30.9\%$ became $-10.0\%$, qwen3-coder's $-18.3\%$ became $-44.0\%$, and gemma3's $-11.7\%$ became $+1.7\%$. 

\begin{table*}[!t]
\caption{Main experiment, local models, before and after the context correction (task level, 36 tasks): aggregate change in output tokens and difference in correct runs under lenient scoring, with Holm-adjusted $p$-values ($p_H$). Bold with an asterisk marks evidence to reject the null hypothesis of no difference, underlined where the treatment was correct more often.}
\label{tab:truncation}
\centering
\begin{fittable}
\begin{tabular}{lrrrrrr}
\toprule
 & \multicolumn{2}{c}{Output tokens, change (\%)} & \multicolumn{2}{c}{Correct, original} & \multicolumn{2}{c}{Correct, corrected} \\
\cmidrule(lr){2-3}\cmidrule(lr){4-5}\cmidrule(lr){6-7}
Model & Original & Corrected & Points & $p_H$ & Points & $p_H$ \\
\midrule
qwen3-coder 30B & $-18.3$ & $-44.0$ & $+7.3$ & 0.31 & \good{$+25.9$} & \sigp{0.0062} \\
qwen2.5-coder 14B & $+30.9$ & $-10.0$ & $+6.0$ & 0.63 & \good{$+30.6$} & \sigp{0.0002} \\
gemma4 26B & $-2.4$ & $-12.9$ & $+0.9$ & 1 & \good{$+20.6$} & \sigp{0.0004} \\
llama3.1 8B & $+6.0$ & $+1.4$ & \good{$+5.1$} & \sigp{0.029} & \good{$+9.2$} & \sigp{0.0002} \\
gemma3 27B & $-11.7$ & $+1.7$ & $-0.9$ & 1 & $-0.9$ & 1 \\
\bottomrule
\end{tabular}
\end{fittable}
\end{table*}

\section{The Function-Patch Contract: Anchoring with Tool-Side Edit Application}
\label{app:patch}

The main experiment's large-tier contract asks the model to decide and apply its own edits. This appendix reports an exploratory experiment that moves the application step into the tooling. Under the \textbf{function-patch contract}, the prompt instructs the model to return only the complete, corrected definitions of the functions it changed. The harness then replaces the same-named functions in the seed, using base R's parser, and validates the resulting program (Fig.~\ref{fig:patch}). A patch that fails to parse, or matches no seed function, is an execution error. This contract stands for a tool-mediated FileMark workflow, in which an editor integration applies the returned fixes. It ran on the same 18 large tasks, 13 replicates, and seven models as the main experiment, and its damaged local cells were re-run with the full context window like the main experiment's.

\begin{figure*}[!t]
  \centering
  \begin{adjustbox}{max width=\textwidth}
\begin{tikzpicture}[
    font=\small\bfseries,
    >={Stealth[length=2.4mm]},
    node distance=4mm and 5mm,
    box/.style={draw, line width=0.5pt, rounded corners=2pt, align=center,
                inner sep=3pt, minimum height=14mm, fill=gray!5, text=black,
                text width=27mm},
    arr/.style={->, line width=1pt, black!70}
  ]
  \node[box, fill=blue!6] (seed) {seed: $K$ named\\functions $+$ driver\\(4 flawed)};
  \node[box, right=of seed, fill=green!8, text width=33mm] (model) {model gets seed $+$\\feedback, returns only\\the corrected functions};
  \node[box, right=of model, fill=orange!8] (splice) {harness replaces\\same-named\\functions (R parser)};
  \node[box, right=of splice, fill=red!6] (val) {full program\\executed and\\validated};
  \draw[arr] (seed) -- (model);
  \draw[arr] (model) -- (splice);
  \draw[arr] (splice) -- (val);
\end{tikzpicture}\end{adjustbox}
  \caption{The function-patch contract. The model returns only the corrected functions, which the harness applies to the seed by name before validation.}
  \label{fig:patch}
\end{figure*}

\textbf{Every local model gains when the tooling applies the edits.} For each model and outcome we test the arm contrast in Hypotheses~\eqref{eq:arm}, as in Sections~\ref{sec:tokens} and~\ref{sec:correctness},
where $\theta_m$ is the mean over the 18 large tasks of the task-level difference between treatment and control under the function-patch contract. The outcomes are the log output tokens and the share of correct runs. Table~\ref{tab:patch} reports the token effect as the aggregate change, as the pre-registered analysis of this wave did, and its $p$-value is the sign-flip test of the task-level log differences. All five local models were correct far more often under the treatment, by 11.5 to 75.6 points, and each has evidence to reject $H_0$ under both scorings (Holm $p \le 0.006$, Table~\ref{tab:patch}). 
gemma3 and llama3.1, which almost never produce a valid edit block, were correct in nearly every treatment run under this contract. The Claude models were correct in every run of both arms and generated 39.3\% (Opus) and 55.5\% (Sonnet) fewer output tokens. Under the edit-block contract on the same tasks they saved 34.0\% and 77.4\%, so the savings do not depend on who applies the edits. 

\begin{table*}[!t]
\caption{Function-patch wave (exploratory), task level (T4--T6, 18 tasks): treatment minus control with 95\% cluster-bootstrap intervals and Holm-adjusted $p$-values, correctness under lenient scoring, and the tasks on which the treatment was more (higher) or less (lower) often correct. Bold with an asterisk marks evidence to reject the null hypothesis of no difference, underlined where the treatment did better.}
\label{tab:patch}
\centering
\begin{fittable}
\begin{tabular}{lrrrrrr}
\toprule
 & \multicolumn{2}{c}{Output tokens} & \multicolumn{4}{c}{Correct runs} \\
\cmidrule(lr){2-3}\cmidrule(lr){4-7}
Model & \hd{Change, \%}{(95\% CI)} & $p_H$ & \hd{Points}{(95\% CI)} & $p_H$ & \hd{Tasks}{higher} & \hd{Tasks}{lower} \\
\midrule
Claude Opus & \goodci{$-39.3$}{$-43.2$}{$-34.8$} & \sigp{$<10^{-4}$} & \ci{0.0}{0.0}{0.0} & n/a & 0 & 0 \\
Claude Sonnet & \goodci{$-55.5$}{$-61.1$}{$-49.4$} & \sigp{$<10^{-4}$} & \ci{0.0}{0.0}{0.0} & n/a & 0 & 0 \\
\addlinespace
qwen3-coder 30B & \ci{$-1.5$}{$-17.5$}{$+15.7$} & 0.57 & \goodci{$+69.2$}{$+48.3$}{$+87.6$} & \sigp{0.0007} & 13 & 0 \\
qwen2.5-coder 14B & \sigci{$+14.7$}{$+3.2$}{$+29.5$} & \sigp{0.047} & \goodci{$+73.9$}{$+58.5$}{$+88.5$} & \sigp{0.0001} & 17 & 0 \\
gemma4 26B & \goodci{$-22.2$}{$-29.8$}{$-13.1$} & \sigp{0.004} & \goodci{$+11.5$}{$+5.1$}{$+19.2$} & \sigp{0.006} & 10 & 1 \\
llama3.1 8B & \goodci{$-53.0$}{$-70.9$}{$-14.5$} & \sigp{$<10^{-4}$} & \goodci{$+75.6$}{$+59.4$}{$+90.6$} & \sigp{0.0001} & 17 & 0 \\
gemma3 27B & \goodci{$-65.5$}{$-79.0$}{$-28.3$} & \sigp{0.041} & \goodci{$+63.2$}{$+42.3$}{$+82.5$} & \sigp{0.0007} & 13 & 0 \\
\bottomrule
\end{tabular}
\end{fittable}
\end{table*}

The contract is a specialized solution tied to R: the harness can splice edits only because our seeds consist of named, top-level function definitions. We therefore report it as exploratory. It shows that anchored feedback helps every local model once the output format stops being the bottleneck, which deserves testing in ecosystems with their own natural edit units, such as methods in Java or functions in Python and Julia.

\textbf{Version sensitivity.} The model-version audit (Section~\ref{sec:models}) left a natural experiment. The 142 cells of this wave first served by \texttt{claude-sonnet-4-6} exist under both Sonnet generations. On these identical cells, the newer Sonnet generated 116.8\% more tokens under the holistic control (398 to 864 mean output tokens) but only 35.0\% more under the anchored treatment (366 to 494). Correctness was at ceiling under both versions (run-level, descriptive). The more verbose generation thus widened the anchoring savings.

\section{Claude Results of the Revision Experiments (Exploratory)}
\label{app:claude_r1}

This appendix asks whether the two Claude models respond to the wording, the format, and a wrong anchor as the local models do (Sections~\ref{sec:decomp} and~\ref{sec:misplaced}). Because Opus was correct in every run and Sonnet in nearly every run, the answer shows mainly in output tokens. The results are kept out of the confirmatory analysis for one reason: their reference arms are three months older. The two Claude models ran every new arm in October 2026, 1{,}044 calls in all. They used the same Claude Code command-line interface (\texttt{claude -p}) as in July, with each model pinned to the identifier its July runs had used. Their control and treatment references are the July runs, made under an older Claude Code version and different instruction files. These results are therefore exploratory, with nominal $p$-values. For each Claude model $m$, outcome, and contrast $k$ they test Hypotheses~\eqref{eq:arm},
where $\theta_{m,k}$ is the mean over the 18 large tasks of the task-level difference of contrast $k$, first arm minus second (Section~\ref{sec:newarms}).

\textbf{Drift check.} Before the new arms, a pre-registered check repeated one replicate of control and treatment on the 18 large tasks for both models (72 calls). A cell would show drift in two cases. Either the median task-level change in output tokens against July left its no-drift range, computed by resampling the July data, and was at least half the model's minimum detectable effect, or two or more tasks failed under lenient scoring. No cell showed drift. The median changes were $-2.5\%$ (Opus control), $-12.0\%$ (Opus treatment), $-10.0\%$ (Sonnet control), and $-2.7\%$ (Sonnet treatment), each inside its range, and no task failed. The check can miss moderate drift, for Sonnet's control up to about 30\%. 

\textbf{Results.} Table~\ref{tab:claude_r1} gives the contrasts. The format at fixed wording cut the typical task's output tokens by 26.7\% (Opus) and 70.5\% (Sonnet). Every misplaced anchor raised them, most of all the offset anchor, which also cut Sonnet's correctness, while Opus stayed correct in every run. At the nominal level we have evidence to reject $H_0$ for every typical-task token contrast except Sonnet's wording contrast, for which we do not have evidence to reject $H_0$ (nominal $p = 0.19$). 

\begin{table*}[!t]
\caption{Decomposition and misplaced-anchor experiments, Claude models against July references (exploratory, nominal $p$-values): output tokens for the typical task and in aggregate, and Sonnet's correctness in points. Bold with an asterisk marks evidence to reject the null hypothesis of no difference, underlined where the arm with the feature under test did better (Table~\ref{tab:contrasts}).}
\label{tab:claude_r1}
\centering
\begin{fittable}
\begin{tabular}{lrrrrr}
\toprule
 & \multicolumn{2}{c}{Opus output tokens (\%)} & \multicolumn{2}{c}{Sonnet output tokens (\%)} & Sonnet correct \\
\cmidrule(lr){2-3}\cmidrule(lr){4-5}\cmidrule(lr){6-6}
Contrast & Typical & Aggregate & Typical & Aggregate & lenient (points) \\
\midrule
\multicolumn{6}{l}{\emph{Decomposition}} \\
Format & \good{$-26.7$} & \good{$-28.9$} & \good{$-70.5$} & \good{$-76.3$} & 0.0 \\ 
Wording & \good{$-10.1$} & $-7.1$ & $-12.0$ & $-4.5$ & 0.0 \\ 
\midrule
\multicolumn{6}{l}{\emph{Misplaced anchors}} \\
5 lines off & \good{$+373.6$} & \good{$+382.4$} & \good{$+297.9$} & \good{$+468.1$} & \good{$-59.3$} \\ 
Caller line & \good{$+57.7$} & \good{$+57.3$} & \good{$+70.0$} & \good{$+91.0$} & 0.0 \\ 
Return line & \good{$+146.0$} & \good{$+164.7$} & \good{$+150.7$} & \good{$+256.0$} & $-1.9$ \\ 
\bottomrule
\end{tabular}
\end{fittable}
\end{table*}

\textbf{Tool use.} Claude Code's command-line interface lets the model use tools, such as reading or searching a file. In 26 of the 1{,}116 calls (2.3\%, against 0.4\% in July), the model read or searched the task's own script before answering, mostly under misplaced anchors. The session transcripts show that no call touched the change list, the validator, or any other experiment file. Without these calls, every contrast keeps its direction and its nominal decision. The offset anchor's cost to Sonnet's correctness grows from 59.3 to 72.5 points, because the calls that checked the script recovered more often. 

\section{Seed Padding and Tier Composition}
\label{app:padding}

Tier membership is defined by lines of code, but the library generator grows programs in whole-function steps of $K$. When the assembled header, functions, and driver fall short of a tier's minimum line count, the generator inserts inert comment lines (\texttt{\# (analysis library continues)}) into the seed's header region. There they cannot disturb flaw-line locations or function structure (\texttt{R/lib\_templates.R}). This appendix quantifies the padding so that the tier construct can be judged precisely. The numbers are computed from the released seeds by \texttt{R/figures.R} (\texttt{results/stats\_seed\_lines.csv}).

\begin{figure*}[!t]
  \centering
  \includegraphics[width=0.8\textwidth]{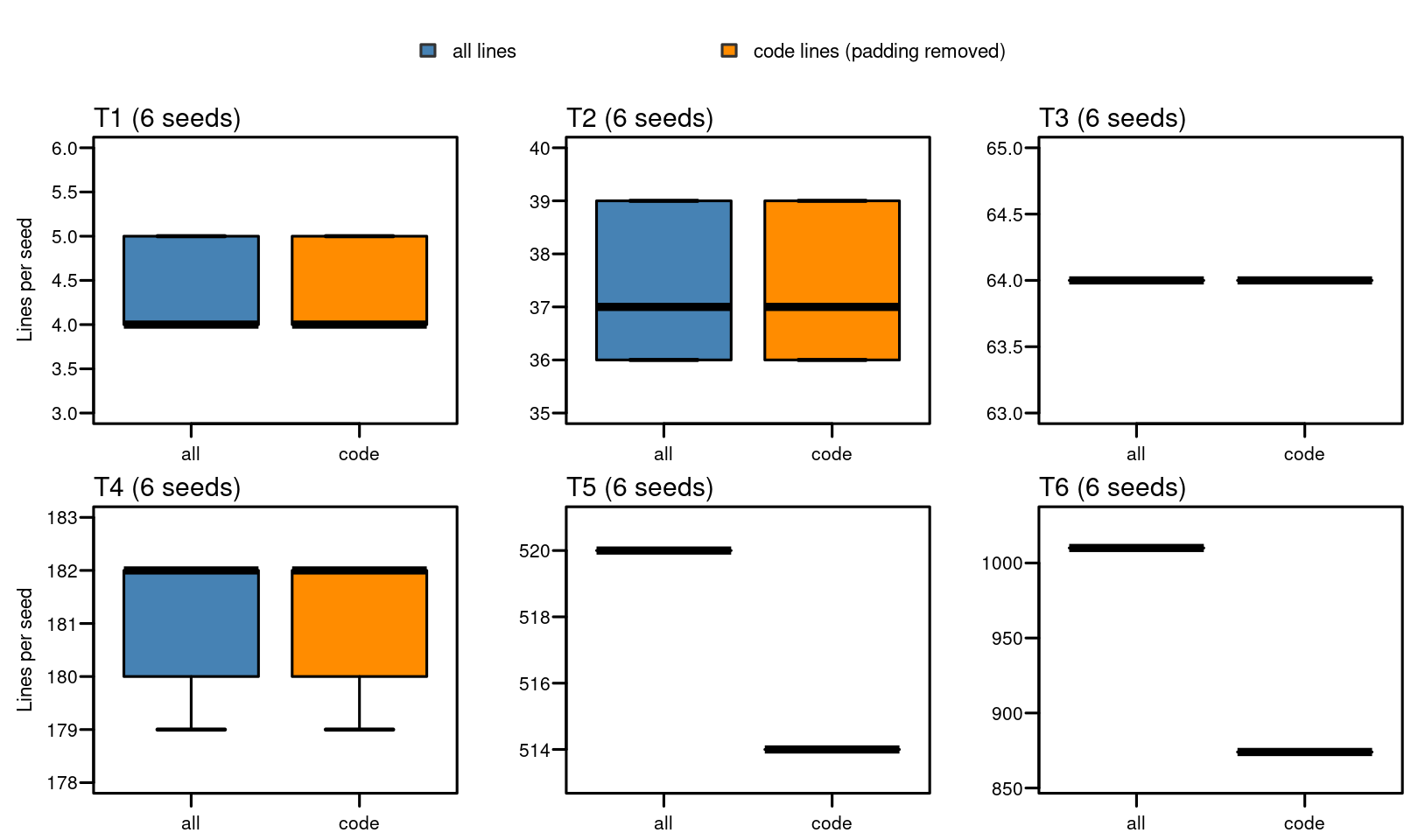}
  \caption{Task bank: seed length by size tier, as boxplots of total and code-only lines over the six seeds of each tier, one panel per tier. Lengths barely vary within a tier, and padding separates the two boxes only on T5 and T6.}
  \label{fig:seed_lines}
\end{figure*}

Fig.~\ref{fig:seed_lines} shows the distribution of seed length per tier, for total lines and for code-only lines (total minus padding), across the six seeds of each tier. Lengths are nearly deterministic by design. Padding is zero for T1--T4, 1.2\% of lines at T5, and 13.5\% at T6, where each seed carries 136 padding lines. The growth between tiers is real analysis code: T5 seeds contain 55 analysis functions and T6 seeds 95 (about 514 and 874 code lines), so the localization search space grows by 73\% from T5 to T6 regardless of padding. T6 crosses its nominal 1{,}000-line boundary only with the padding included, and its code alone would sit at the top of T5's range. The padding is one contiguous, trivially skippable block, identical in both arms of every pair, and it adds only to raw file length, so it cannot confound the paired treatment effect. The effective difficulty driver on large tiers is the number of candidate functions a model must search.

\subsection*{Are the six tasks in a tier the same question?}

Are a tier's six tasks effectively one question repeated, which would overstate the per-tier sample size? We audited this directly (\texttt{R/audit\_task\_distinctness.R}). A \emph{question} is the pair of seed program and requested change set. We measured exact identity with MD5 hashes and structural overlap with the Jaccard index of padding-free code lines \cite{jaccard_distribution_1912}. \textbf{All 36 seeds are distinct} (36 of 36 unique hashes), so no two tasks pose the same question. Each tier's six tasks also carry six distinct themes (Table~\ref{tab:distinct}). T6, for instance, spans sensor calibration, a clinical trial, a marketing funnel, sales forecasting, sensor drift, and genomics quality control. The requested change sets are 35 of 36 distinct. The one exception is a T2 pair (\texttt{t2\_lib\_01} and \texttt{t2\_lib\_06}) that applies the same two-edit instruction set to two different seed programs, so even those two remain different questions.

What the tasks share is deliberate. The tasks are variants of one parameterized family, so within a tier they reuse a common skeleton: the data generator, the driver, and block types from a fixed pool. The result is substantial line-level overlap, with mean pairwise Jaccard similarity of 0.57 to 0.80 on tiers T2--T6 and near zero only on the tiny T1 files. The most similar pairs reach 0.86 to 0.94. These near-duplicate seeds differ mainly in the injected flaw and the dataset theme (Table~\ref{tab:distinct}, ``Jaccard max''). The injected flaws also come from a finite taxonomy: omitted predictor, missing \texttt{na.rm}, wrong summary statistic, wrong standardization denominator, and wrong threshold. The same type of flaw, and often the same instruction text, therefore recurs across tasks, and only 26 of the 129 feedback strings are distinct (20\%). Each instance is still bound to a different function, line, and themed dataset. On the short tiers a few tasks share a set of flawed line positions (Table~\ref{tab:distinct}, T1--T3), because a short file has few candidate lines. This uniformity of structure with variation of content lets file size stand alone as the tier construct. The tasks are distinct problems, but they are deliberately near-relatives rather than independent draws. The intended repetition in the design is the 13 replicates per task, which re-ask the same question and are averaged within each task in the analysis (Section~\ref{sec:power}).

\begin{table*}[!t]
\caption{Task bank: within-tier task distinctness (six tasks per tier, from \texttt{R/audit\_task\_distinctness.R}). Columns count distinct seeds, requested change sets, themes, and sets of flawed line numbers, and give the mean and maximum pairwise Jaccard similarity of padding-free code lines.}
\label{tab:distinct}
\centering
\begin{fittable}
\begin{tabular}{lcccccc}
\toprule
Tier & Seeds & Change sets & Themes & Flaw lines & Jaccard mean & Jaccard max \\
\midrule
T1 & 6/6 & 6/6 & 6/6 & 3/6 & 0.06 & 0.29 \\
T2 & 6/6 & 5/6 & 6/6 & 3/6 & 0.57 & 0.94 \\
T3 & 6/6 & 6/6 & 6/6 & 5/6 & 0.80 & 0.89 \\
\rowcolor{tablegray} T4 & 6/6 & 6/6 & 6/6 & 6/6 & 0.67 & 0.86 \\
\rowcolor{tablegray} T5 & 6/6 & 6/6 & 6/6 & 6/6 & 0.59 & 0.91 \\
\rowcolor{tablegray} T6 & 6/6 & 6/6 & 6/6 & 6/6 & 0.57 & 0.93 \\
\bottomrule
\end{tabular}
\end{fittable}
\end{table*}

\section{Complete Prompts and Localization Cues}
\label{app:prompts}
The original study sent 108 distinct prompts, 72 of them in the main experiment, and the revision experiments added 174, for 282 in all. A prompt depends only on the task, the arm, and the output contract, never on the model or the replicate, and all of them are in the repository (\texttt{tasks/} and each run's \texttt{prompt.txt}). This appendix prints one pair in full: task \texttt{t2\_lib\_01} under the full-rewrite contract. Prompt lines 1--51 are identical in both arms. The two prompts differ only from line 52 on. As in Fig.~\ref{fig:prompts}, a hooked arrow ($\hookrightarrow$) at the start of a printed line marks a long line wrapped to fit the page. The arrow is a typesetting aid, not a character of the prompt.

\medskip\noindent\textbf{Shared prompt lines 1--51 (both arms).}
\begin{Verbatim}[fontsize=\scriptsize, frame=single, breaklines]
You are an expert R programmer. You are given an R script that is flawed or incomplete, followed by requested changes.

Requirements:
- use base R only: no library() or require() calls, and no external packages (no ggplot2, dplyr, tidyverse, data.table, caret, etc.)
- preserve the variable and function names referenced in the requests
- Return the COMPLETE corrected script (every line), ready to run with Rscript.

Respond with ONLY the R code in a single ```r code block, and nothing else.

--- Current script (seed.R) ---
# Analysis library: sensor calibration
# Base R only. make_data() builds the dataset; analyze_block_* compute results;
# the driver at the bottom runs every block and stores its result in res_<i>.

make_data <- function(n = 150) {
  x1 <- rnorm(n)
  x2 <- runif(n, -1, 1)
  x3 <- rnorm(n, 5, 2)
  x3[seq(1, n, by = 10)] <- NA            # inject missingness
  x4 <- as.numeric(rpois(n, 3))
  g  <- factor(sample(c('a','b','c'), n, replace = TRUE))
  y  <- 1.5 * x1 - 0.8 * x2 + 0.3 * x4 + rnorm(n)
  data.frame(y, x1, x2, x3, x4, g)
}

analyze_block_1 <- function(d) {
  # two-predictor linear model and its in-sample RMSE
  m <- lm(y ~ x1, data = d)            # [block 1]
  preds <- predict(m)
  resid <- d$y - preds
  rmse  <- sqrt(mean(resid^2))
  r2    <- summary(m)$r.squared
  list(rmse = rmse, r2 = r2, n = nrow(d))
}

analyze_block_2 <- function(d) {
  # average of x3, which contains missing values
  mu <- mean(d$x3)                     # [block 2]
  sdv <- sd(d$x3, na.rm = TRUE)
  rng <- range(d$x3, na.rm = TRUE)
  list(mu = mu, sdv = sdv, rng = rng)
}


# ---- driver ----
set.seed(42)
d <- make_data()
res_1 <- analyze_block_1(d)
res_2 <- analyze_block_2(d)

--- Requested changes ---
\end{Verbatim}

\noindent\begin{minipage}[t]{0.40\linewidth}
  \footnotesize\textbf{Control, lines 52--56}
  \begin{Verbatim}[fontsize=\scriptsize, frame=single, breaklines]
Please make the following changes to the script:

1. Include x2 in the linear model (y ~ x1 + x2) for this block.
2. Compute the mean of x3 with na.rm = TRUE so it is not NA.

  \end{Verbatim}
\end{minipage}\hfill
\begin{minipage}[t]{0.58\linewidth}
  \footnotesize\textbf{Treatment, lines 52--80}
  \begin{Verbatim}[fontsize=\scriptsize, frame=single, breaklines, breakafter=/, breakaftersymbolpre={}]
# Feedback: seed.R
File: `/home/billy/Documents/Projects/research/filemark_usage/tasks/t2_lib_01/seed.R`
Exported: deterministic (generated)

---

## Comment 1 (line 18)

**Selected text:**
```
m <- lm(y ~ x1, data = d)            # [block 1]
```

**Feedback:** This model omits predictor x2. Change it to lm(y ~ x1 + x2, data = d) so the fit uses both predictors.

---

## Comment 2 (line 28)

**Selected text:**
```
mu <- mean(d$x3)                     # [block 2]
```

**Feedback:** x3 contains NA, so mean(d$x3) is NA. Use mean(d$x3, na.rm = TRUE) so mu is finite.

---


  \end{Verbatim}
\end{minipage}

\medskip
For the task of Fig.~\ref{fig:prompts}, \texttt{t4\_lib\_01}, the change list both arms are rendered from is reproduced below. Its 200 shared prompt lines are in the repository (\url{results/runs/claude_sonnet__t4_lib_01__rep01__control__editblock/prompt.txt}).

\begin{Verbatim}[fontsize=\scriptsize, frame=single, breaklines]
{
  "task_id": "t4_lib_01",
  "template_id": "t4_lib",
  "tier": "T4",
  "task_type": "library",
  "plotting": false,
  "output_mode": "patch",
  "n_seed_lines": 182,
  "task_instruction": "",
  "changes": [
    {
      "start_line": 36,
      "end_line": 36,
      "feedback": "This should report group MEANS, not medians. Use tapply(d$y, d$g, mean).",
      "control_item": "Compute group means of y by g (mean, not median).",
      "selected_text": "gm <- tapply(d$y, d$g, median)       # [block 3]"
    },
    {
      "start_line": 68,
      "end_line": 68,
      "feedback": "x3 contains NA, so mean(d$x3) is NA. Use mean(d$x3, na.rm = TRUE) so mu is finite.",
      "control_item": "Compute the mean of x3 with na.rm = TRUE so it is not NA.",
      "selected_text": "mu <- mean(d$x3)                     # [block 7]"
    },
    {
      "start_line": 131,
      "end_line": 131,
      "feedback": "The threshold is wrong: count x1 > 1 (not x1 > 0). Use sum(d$x1 > 1).",
      "control_item": "Count observations with x1 > 1 (not > 0).",
      "selected_text": "cnt <- sum(d$x1 > 0)                 # [block 15]"
    },
    {
      "start_line": 138,
      "end_line": 138,
      "feedback": "This model omits predictor x2. Change it to lm(y ~ x1 + x2, data = d) so the fit uses both predictors.",
      "control_item": "Include x2 in the linear model (y ~ x1 + x2) for this block.",
      "selected_text": "m <- lm(y ~ x1, data = d)            # [block 16]"
    }
  ]
}
\end{Verbatim}

Table~\ref{tab:cues} counts, over all 129 change items, the cues each arm's text carries. The counts come from regular expressions over the texts in \texttt{changes.json}, with the matched evidence saved per item (\texttt{results/content\_audit\_items.csv}) and checked by hand. The ``as sent'' column adds what the FileMark comment carries around the treatment text: a line label, the quoted line, and, from T2 on, the \texttt{[block N]} tag inside that line. In 78 of the 110 single-line items the flawed line has look-alike lines in its script (same assigned name and called function, median four), which a search by content must tell apart.

\begin{table*}[!t]
\caption{Task bank: localization cues in the 129 change items, by arm.}
\label{tab:cues}
\centering
\begin{fittable}
\begin{tabular}{lrrr}
\toprule
 & Control & \multicolumn{2}{c}{Treatment} \\
\cmidrule(lr){3-4}
Cue & text & text & as sent \\
\midrule
Line number & 0 & 0 & 129 \\
Function name or block tag & 0 & 0 & 108 \\
Quoted source line & 0 & 0 & 129 \\
Exact replacement code & 1 & 112 & 112 \\
Any code fragment & 98 & 115 & 115 \\
Script variable name & 115 & 115 & 115 \\
Error description (diagnosis or ``not X'') & 65 & 113 & 113 \\
Wording that presupposes an anchor & 24 & 42 & 42 \\
\midrule
Mean length (characters) & 55 & 90 & 90 \\
\bottomrule
\end{tabular}
\end{fittable}
\end{table*}

\section{Power and Clustering}
\label{app:power}
The original calculation, made before the full runs, used a pilot of 42 pairs (one task per tier, one replicate, seven models) and sized pairs, not tasks. Its targets were a two-sided test at $\alpha = 0.05$ with power 0.80. The pilot's paired score effect (0.119, standard deviation 0.309) gave 55 pairs per model in the paired-sample formula of \texttt{power.t.test}, and its correctness proportions 0.595 and 0.786 gave 92 per arm in the two-sample \texttt{power.prop.test}. Both treat pairs as independent, the first also assumes normal differences, and the pilot's local large-tier prompts were truncated like the full runs. The 36 tasks were an input that the calculation converted into replicates per task. Its recommendation was two replicates (72 pairs per model), which met the score requirement but not the correctness one. The 13 replicates run (468 pairs) were a choice above both. With 13 replicates the design effect $1 + 12\rho$ for the paired differences ranges from 2.4 to 12.4 across models and outcomes, where $\rho$ is the intraclass correlation within a task. The unit of inference is now the task, so the planned power does not apply to the tests reported here. Their power rests on the 36 tasks, which no calculation sized. More tasks, not more replicates, would raise it. We therefore report the smallest effects the design could detect, from the observed variability, rather than post hoc power, which adds nothing to the observed $p$-values \cite{hoenig_abuse_2001}.

Table~\ref{tab:power} gives, per model, the clustering of the paired differences and the minimum detectable effect (MDE) at the task level for a two-sided test at $\alpha = 0.05$ with power 0.80 (Appendix~\ref{app:stats}). Its standard deviation is the one observed in the main experiment, where the MDE is 4\% to 40\% for output tokens and 3 to 21 points for correctness, depending on the model. An effect that varies more from task to task must be larger to be detected. 
For an effect already tested, the comparison with its MDE restates its $p$-value. An estimate equals its MDE when its $t$ statistic is 2.88 for 36 tasks or 2.97 for 18 (two-sided $p$ 0.0067 or 0.0085). The MDEs therefore add nothing to the established effects, all of which exceed them (Table~\ref{tab:power}), and matter for the contrasts without evidence. 
The new arms compare 3 replicates with the reference's 13 on only 18 tasks, so they can detect only larger effects, up to 78\% for tokens and 35 points for correctness. 
A single tier has six tasks, so its sign-flip $p$ is 0.031 at best and its Holm-adjusted $p$ over six tiers 0.19 at best (Section~\ref{sec:safeguards}). Its MDE is 1.42 standard deviations of the tier's task-level differences, for example 16.7\% of output tokens for Opus and 34.7 points of correctness for qwen2.5-coder, which is why per-tier results are descriptive. 
The conclusion is asymmetric: a new-arm contrast with evidence to reject the null hypothesis shows an effect, but one without that evidence is not evidence of no effect. The wording effect of $+27.6$ points for qwen3-coder (Section~\ref{sec:decomp}), for example, is below that model's new-arm MDE of 34.9 points, so an effect of that size can neither be shown nor ruled out. Such contrasts are reported as estimates with intervals, not as findings, and Holm correction raises the detectable effects further. 

\begin{table*}[!t]
\caption{Clustering and minimum detectable effects (MDE) per model, from the observed standard deviation of the task-level differences, with correctness under lenient scoring. The main experiment compares 13-replicate arms on 36 tasks, and a new arm compares 3 replicates with a 13-replicate reference on 18 tasks.}
\label{tab:power}
\centering
\begin{fittable}
\begin{tabular}{lrrrrrrrr}
\toprule
 & \multicolumn{3}{c}{Output tokens, main experiment} & \multicolumn{3}{c}{Correctness, main experiment} & \multicolumn{2}{c}{New arm MDE} \\
\cmidrule(lr){2-4}\cmidrule(lr){5-7}\cmidrule(lr){8-9}
Model & ICC & Design effect & MDE & ICC & Design effect & MDE & Tokens & Correctness \\
\midrule
Claude Opus & 0.44 & 6.3 & 14.8\% & n/a & --- & ceiling & 14.3\% & ceiling \\
Claude Sonnet & 0.78 & 10.3 & 39.6\% & 0.14 & 2.7 & 2.5 & 28.8\% & 11.3 \\
\addlinespace
qwen3-coder 30B & 0.61 & 8.3 & 29.4\% & 0.95 & 12.4 & 20.8 & 69.5\% & 34.9 \\
qwen2.5-coder 14B & 0.60 & 8.3 & 14.9\% & 0.67 & 9.0 & 19.3 & 32.7\% & 28.3 \\
gemma4 26B & 0.26 & 4.2 & 11.9\% & 0.36 & 5.4 & 14.2 & 11.5\% & 24.9 \\
llama3.1 8B & 0.30 & 4.7 & 28.5\% & 0.11 & 2.4 & 6.8 & 77.8\% & 12.2 \\
gemma3 27B & 0.72 & 9.6 & 4.3\% & 0.14 & 2.7 & 2.5 & 3.4\% & floor \\
\bottomrule
\end{tabular}
\end{fittable}
\end{table*}

\begin{figure*}[!t]
  \centering
  \begin{adjustbox}{max width=0.75\linewidth}
\begin{tikzpicture}[
    font=\footnotesize,
    >={Stealth[length=2.2mm]},
    ev/.style={draw, line width=0.4pt, rounded corners=2pt, fill=gray!6, align=left,
               text width=76mm, inner sep=2.5pt, anchor=west},
    loc/.style={align=left, text width=30mm, anchor=west},
    day/.style={anchor=east, align=right, font=\footnotesize\bfseries},
    phase/.style={anchor=west, font=\small\bfseries}
  ]
  \draw[line width=1pt, black!55, ->] (0, 0.15) -- (0, -11.3);
  \node[phase] at (-2.0, 0.5) {Original study, June and July 2026};
  \node[anchor=west, font=\footnotesize\itshape] at (8.45, 0.5) {Results in};
  \newcommand{\event}[4]{%
    \fill[black!70] (0, #1) circle (2.2pt);
    \node[day] at (-0.2, #1) {#2};
    \node[ev] at (0.25, #1) {#3};
    \node[loc] at (8.45, #1) {#4};}
  \event{-0.45}{Jun 25}{Pilot: 42 pairs (one task per tier, one replicate, seven models) set the sample size.}{Section~\ref{sec:power}}
  \event{-1.45}{Jun 25--Jul 5}{Main experiment on T1--T3 (full rewrites) and the function-patch wave on T4--T6, 3{,}276 runs each.}{Sections~\ref{sec:tokens}--\ref{sec:interaction}, Appendix~\ref{app:patch}}
  \event{-2.45}{Jul 6--7}{Main experiment on T4--T6 (edit blocks), 3{,}276 runs, completing its 6{,}552 runs.}{Sections~\ref{sec:tokens}--\ref{sec:failure}}
  \event{-3.45}{July}{First analysis: run-level tests on truncated local prompts and first-block scoring.}{Superseded, Appendix~\ref{app:truncation}}
  \node[phase] at (-2.0, -4.4) {Revision, October 2026};
  \event{-5.35}{Oct 9}{Plans committed, truncation found, and all 9{,}828 runs re-scored under both scorings.}{Sections~\ref{sec:outcomes}--\ref{sec:power}}
  \event{-6.35}{Oct 9--12}{Local re-runs with a 32{,}768-token window, 4{,}836 runs: the corrected data.}{All local results, Appendix~\ref{app:truncation}}
  \event{-7.35}{Oct 9--14}{Hosted drift check (72 calls, passed Oct 10) and hosted new arms (1{,}044 calls).}{Appendix~\ref{app:claude_r1}}
  \event{-8.35}{Oct 12--13}{Local new arms: decomposition, misplaced anchors, and contract control, 2{,}610 runs.}{Sections~\ref{sec:decomp}--\ref{sec:contractctl}}
  \event{-9.35}{Oct 14}{Pre-registered task-level analysis of all runs.}{Section~\ref{sec:results}, Appendices~\ref{app:patch}, \ref{app:tiers}, \ref{app:power}}
  \event{-10.4}{Oct 22--31}{Checks added after the results: hierarchical models, interaction tests, both token estimands, and scale and tokenizer checks.}{Section~\ref{sec:power}, Appendices~\ref{app:splitplot}, \ref{app:hier}}
  \node[anchor=north, font=\footnotesize\itshape, black!60] at (0, -11.35) {time};
\end{tikzpicture}\end{adjustbox}
  \caption{Timeline of all runs and analyses, main experiment and revision experiments, with the section or appendix that reports each. The corrected data are the October re-runs of the damaged local cells together with the undamaged original runs.}
  \label{fig:timeline}
\end{figure*}

\section{Execution Sandbox}
\label{app:sandbox}
Every generated script is executed and validated by the same procedure in both arms (\texttt{R/run\_candidate.R}, \texttt{R/score.R}, and \texttt{R/config.R} in the repository), under the following settings.
\begin{itemize}
  \item Process: each script runs in its own new R process (\texttt{Rscript}), so a crash or a hang cannot affect the harness or another run.
  \item Working directory: the run's own directory, where any file the script writes, such as a required plot, is created and later checked.
  \item Environment: the script is evaluated in a fresh R environment, separate from the validator's code, and warnings are muffled, so that only errors stop it.
  \item Time limit: 30 seconds of CPU time and 30 seconds of elapsed time, enforced by R's \texttt{setTimeLimit}. A script that exceeds either limit is an execution error.
  \item Validation: after the script finishes, the limit is lifted, and the task's validator inspects the objects the script created and recomputes the expected values from them.
  \item Packages: a static check of the script flags any package other than those that ship with R (constraint adherence, Section~\ref{sec:outcomes}).
\end{itemize}
The sandbox separates runs and bounds their running time, but it is not a security boundary: it uses no container or virtual machine and does not restrict network or file-system access. Every run used R 4.5.2 on the workstation of Section~\ref{sec:env}, and no script in either arm used a package outside base R (Section~\ref{sec:failure}).

\section{Timeline of Runs and Analyses}
\label{app:timeline}
Figure~\ref{fig:timeline} orders the runs and analyses, from the June 2026 pilot to the October 2026 re-runs and new arms, and names where each result appears.

\end{document}